# Application of Self-Organizing Maps for clustering DJIA and NASDAQ100 portfolios

A.A.Zherebtsov[*], Yu.A.Kuperin[#]

[*]**Division of Computational Physics, Department of Physics,
St.Petersburg State University, St.Petersburg 198904, Russia
e-mail: jerebtsov@AZ9629.spb.edu**
[#]**Laboratory of Complex Systems Theory, Institute of Physics,
St.Petersburg State University, St.Petersburg 198904, Russia
e-mail: kuperin@JK1454.spb.edu**

## Abstract

In this paper we apply the Self -Organized Map (SOM) method for clustering the DJIA and NASDAQ100 portfolios for determination of non-linear correlations between stocks. We represent the application of this method as alternative to ultrametric spaces method. We have found , that SOM method is more relevant and perspective for clustering ill-structured large databases and, in particular, NASDAQ100, where nonlinear processing of the large data samples is required.

## 1 Introduction

Usually the process of data analysis or extraction of knowledge from the data consists of a number of iterative steps, since the formulation of the purposes depends in some respects on the obtained results. It can include a loop of a feedback that means reformulation the purposes on the basis of the received information. Depending on the purposes and complexity of the data it is possible to use any type of well-known algorithms based on recognition of images, machine training or the multivariate statistical analysis. The key point here is the detection of originally unknown structures or patterns in the analyzed data.

The most popular and simple approach to generalization of data sets is statistical tables. Elementary of them allows to obtain the statistics of the data. It could be, for example, the minimum and maximum values in a data set, a median, the first and the third quartile etc. It works well for linear data sets of small dimension. However it remains the rather important combined practical problem of generalization and visualization of the multidimensional data sets.

The Self-Organizing Map represents the result of a vector quantization algorithm that places a number of reference or codebook vectors into a high-dimensional input data space to approximate to its data sets in an order fashion (Kohonen, 1982,1990,1995, Kohonen, Oja, et al, 1996). When local-order relations are defined between the reference vectors, the relative values of the latter are made to depend on each other as if their neighboring values would lie along an "elastic surface". By means of the self-organizing algorithm, this "surface" becomes defined as a kind of nonlinear regression of the reference vectors through the data points. A mapping from a high-dimensional data space into a two-dimensional lattice of points is thereby also defined. Such a mapping can effectively be used to visualize metric ordering relations of input samples. The self-organized map algorithm has been used for a wide variety of applications, mostly for engineering problems but also for data analysis (Back, et al (1996), Demartines (1994), Carlson (1991), Cheng, et al (1994), Garavaglia (1993), Martín-del-Brío, Serrano-Cinca (1993), Marttinen (1993), Serrano-Cinca (1996), Ultsch (1993), Ultsch, et al (1990), Varfis, et al (1992), Zhang, Li, (1993), Deboeck, Kohonen (1998))).

In Mantegna, Stanley (2001) the method of ultrametric spaces was used for clustering of DJIA and NASDAQ500 portfolios. In the present paper we apply both the ultrametric spaces approach and SOM algorithms for clustering the DJIA and NASDAQ100. We compare these approaches and show, that in contrast to the ultrametric spaces method the SOM algorithms are more perspective for clustering NASDAQ100 where processing of the large data samples is required. It is due to the linear character of methods of hierarchical trees and intrinsically nonlinear processing of information by any neural network and by SOM, in particularly. Finally SOM can learn from the data, i.e. it is adaptive algorithm for data clustering or quantization whereas the ultrametric spaces method is not.

The paper is organized as follows. In Sections II and III, we review the basics principles of SOM and ultrametric spaces methods accordingly. In Section IV, we present the study of the DJIA index using ultrametric spaces method. In Section V, clusterization of DJIA by SOM method is represented. Sections VI and VII, presents clusterization of NASDAQ100 by ultrametric spaces and SOM methods accordingly. In Section VIII some conclusions are given.

## 2    The principle of SOM

SOM is a kind of neural network learning without a supervisor (Kohonen, 1982,1990,1995, Kohonen, Oja, et al, 1996). Adapting to the training set SOM forms its outputs by itself. The basic SOM consists of M neurons located on a regular low-dimensional lattice, usually 1- or 2-dimensional. Higher dimensional lattices are possible, but they are not generally used since their visualization is problematic. The lattice is usually either hexagonal or rectangular.

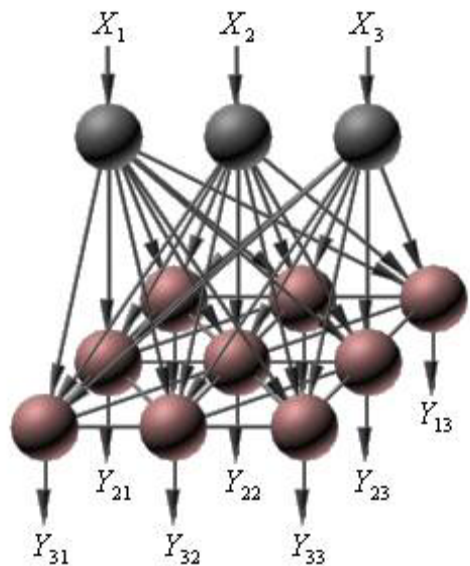

Figure 1.

The schematic architecture of Self-Organized Map.
Gray balls form the input layer, and brown ones form the output layer. Arrows connecting
input and output layers are connections with its own weights.
(After Shumsky et al (1998)).

It is well-known (Kohonen (1995)) that SOM performs two types of data compression:

- reduction of data dimension with minimum lost of information. (This neural networks can single out sets of independent characteristics);

- reduction of data variety due to terminal composition prototypes separation. (Clustering and quantization of data sets);

The basic SOM algorithm is iterative. Each neuron $i$ has a $d$-dimensional prototype vector $\mathbf{w}_i = [w_{i1},...,w_{id}]$ or weight of $i$-th neuron. At each training step, a sample data vector $x$ is randomly chosen from the training set. Distances between $x$ and all the prototype vectors are computed. The best-matching unit (BMU) or the winner unit, denoted here by $\mathbf{x}_{i^*}$, is the map unit with prototype closest to $x$ (Kaski (1997)):

$$|\mathbf{w}_{i^*} - \mathbf{x}| \leq |\mathbf{w}_i - \mathbf{x}|, \forall i \neq i^*.$$

Next, the prototype vectors are updated. The BMU and its topological neighbors are moved closer to the input vector in the input space by the rule $\Delta \mathbf{w}_{i^*}^r = \eta(\mathbf{x}^r - \mathbf{w}_{i^*})$, where. $\eta$ is learning rate and $\Delta \mathbf{w}_{i^*}^r$ is modification of $i$-th neuron weight.

Finally, the update rule for the all vectors of unit $i$ is:

$$\Delta \mathbf{w}_i^\tau = \eta \Lambda(|\mathbf{i} - \mathbf{i}^*|)(\mathbf{x}^\tau - \mathbf{w_i}),$$

where $\Lambda(|\mathbf{i} - \mathbf{i}^*|)$ is a neighborhood kernel centered on a winner unit. The kernel can be for example Gaussian: $\Lambda(a) = \exp(-a^2/\sigma^2)$, where $\sigma$ is neighborhood radius. Both learning rate $\eta$ and neighborhood radius $\sigma$ decrease monotonically in time. During training, the SOM behaves like a flexible net that folds into the "cloud" formed by the training data (Fig. 2). Because of the neighborhood relations, neighboring prototypes are pulled to the same direction, and thus prototype vectors of neighboring units resemble each other. Amount of neurons in output layer defines maximum difference of model vectors. Then trained SOM can classify its inputs: BMU define class of input vectors.

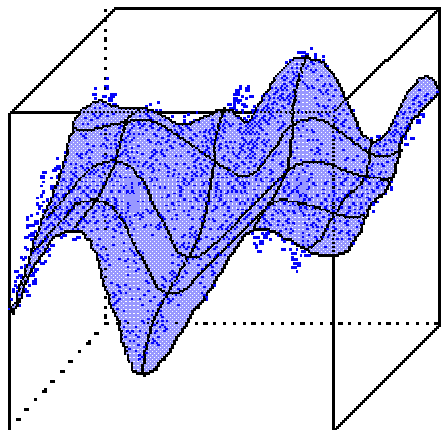

Figure 2.

2-dimensional map of 3-dimensional data set.
(After Shumsky et al (1998)).

The SOM forms a low-dimensional map of training set. The ordered SOM lattice can be used as a convenient visualization platform for showing different features of the SOM (and thus of the data). The goal of visualization is to present large amount of detailed information in order to give a qualitative idea of the properties of the data. Typically the number of properties that need to be visualized is much higher than the number of usable visual dimensions. It is simply impossible to show them all in a single figure. Every vector from many-dimensional input space have it is own coordinate on the lattice. The closer coordinates of two vectors on the map are, the closer these vectors are in the input space. But the opposite statement is not correct (Fig. 3). Representation reducing dimension and retaining nearness attitude does not exist in general case.

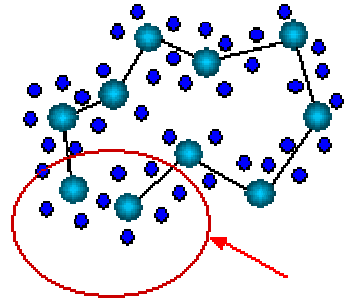

Figure 3.

Close vectors from the input space are reflected on the opposite sides of the map.
(After Shumsky et al (1998)).

It is convenient to visualize SOM like topographic map. Each characteristic of the input data causes its own coloration on the map (Fig. 4) (Kohonen, Oja et al (1996)).

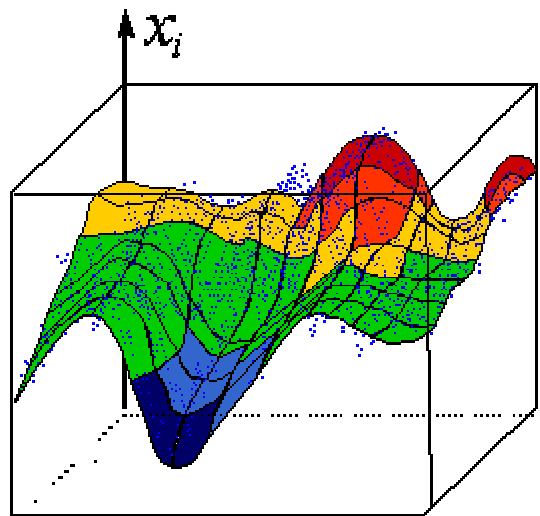

Figure 4.

Coloring of the topographic map induced by i-th component of input data.
(After Shumsky et al (1998)).

# 3 Ultrametric spaces

In financial markets, huge number of stocks is traded simultaneously. One way to detect similarities and differences in the synchronous time evolution of a pair of stocks is to study the correlation coefficient $\rho_{ij}$ between the daily logarithmic changes in price of two stocks $i$ and $j$ (Mantegna, Stanley (2001)).

Let us define logarithmic return for stock $i$

$$S_i(t) = \ln Y_i(t) - \ln Y_i(t-1), \qquad (1)$$

where $Y_i(t)$ is a daily closure price of stock $i$ at time $t$ ($0 \leq t \leq n$). Correlation coefficient $\rho_{ij}$ between the daily logarithmic changes in price of two stocks $i$ and $j$ is given by:

$$\rho_{ij} = \frac{\langle S_i S_j \rangle - \langle S_i \rangle \langle S_j \rangle}{\sqrt{\langle S_i^2 - \langle S_i \rangle^2 \rangle \langle S_j^2 - \langle S_j \rangle^2 \rangle}} \qquad (2)$$

The angular brackets indicate a time average over all the trading days within the investigated time period.

Let

$$\hat{S}_i = \frac{S_i - \langle S_i \rangle}{\sqrt{\langle S_i^2 \rangle - \langle S_i \rangle^2}} \qquad (3)$$

Then we can consider $\hat{S}_i(t)$ ($1 \leq t \leq n$) as the vectors $\vec{\hat{S}}_i$ in n-dimensional space (Rammal et al (1986)).

One can set the distance in this space by several well-known ways:

- Euclidean distance: $d_{ij} = \|\vec{\hat{S}}_i - \vec{\hat{S}}_j\| = \sqrt{\sum_{k=1}^{n}(\hat{S}_{ik} - \hat{S}_{jk})^2}$ ;

- squared Euclidean distance: $d_{ij} = \|\vec{\hat{S}}_i - \vec{\hat{S}}_j\|^2 = \sum_{k=1}^{n}(\hat{S}_{ik} - \hat{S}_{jk})^2$ ;

- City-block (Manhattan) distance: $d_{ij} = \sum_{k=1}^{n}|\hat{S}_{ik} - \hat{S}_{jk}|$ ;

- Chebychev distance: $d_{ij} = \max_k |\hat{S}_{ik} - \hat{S}_{jk}|$ ;

- Power distance: $d_{ij} = (\sum_{k=1}^{n}|\hat{S}_{ik} - \hat{S}_{jk}|^P)^{1/r}$, where r and p are free parameters;

- Pearson distance: $d_{ij} = 1 - \rho_{ij}$, here $\rho_{ij}$ is defined by equation (2).

All distances written above should satisfy the standard properties:
1. $d_{ij} > 0$; $d_{ij} = 0$, если $i = j$;
2. $d_{ij} = d_{ji}, \forall i,j$ ;
3. $d_{ij} \leq d_{ik} + d_{kj}, \forall i,j,k$ ;

An ultrametric space is a space endowed by so-called ultrametric distance. An ultrametric distance $\hat{d}_{ij}$ must satisfy the first two properties of a metric distance, while the usual triangular inequality is replaced by a stronger inequality, called an ultrametric inequality,

$$\hat{d}_{ij} \leq \max_k \{\hat{d}_{ik}, \hat{d}_{kj}\} .$$

Benzécri (1984) rigorously studied the general connection between indexed hierarchies and ultrametrics. Provided that a metric distance between *n* objects exists, several ultrametric spaces can be obtained by performing any given partition of *n* objects. By means of ultrametric spaces one can obtain hierarchical trees.

However, once several objects have been linked together in some clusters, we should determine the distances between those new clusters. In other words, we should determine the rules of amalgamation (linkage rules). Some of this rules adducing below.

- **Single linkage (nearest neighbor)**
  In this method the distance between two clusters is determined by the distance of the two closest objects (nearest neighbors) in the different clusters. This rule will, in a sense, string objects together to form clusters, and the resulting clusters tend to represent long "chains."
- **Complete linkage (furthest neighbor)**
  In this method, the distances between clusters are determined by the greatest distance between any two objects in the different clusters (i.e., by the "furthest neighbors"). This method usually performs quite well in cases when the objects actually form naturally distinct "clumps." If the clusters tend to be somehow elongated or of a "chain" type nature, then this method is inappropriate.
- **Unweighted pair-group average**
  In this method, the distance between two clusters is calculated as the average distance between all pairs of objects in the two different clusters. This method is also very efficient when the objects form natural distinct "clumps," however, it performs equally well with elongated, "chain" type clusters.
- **Weighted pair-group average**
  This method is identical to the unweighted pair-group average method, except that in the computations, the size of the respective clusters (i.e., the number of objects contained in them) is used as a weight. Thus, this method (rather than the previous method) should be used when the cluster sizes are suspected to be greatly uneven.
- **Unweighted pair-group centroid**
  The centroid of a cluster is the average point in the multidimensional space defined by the dimensions. In a sense, it is the center of gravity for the respective cluster. In this method, the distance between two clusters is determined as the difference between centroids.
- **Weighted pair-group centroid (median)**
  This method is identical to the previous one, except that weighting is introduced into the computations to take into consideration differences in cluster sizes (i.e., the number of objects contained in them). Thus, when there are (or one suspects there to be) considerable differences in cluster sizes, this method is preferable to the previous one.
- **Ward's method**
  This method is distinct from all other methods because it uses an analysis of variance approach to evaluate the distances between clusters. In short, this method attempts to minimize the Sum of Squares (SS) of any two (hypothetical) clusters that can be formed at each step. Refer to Ward (1963) for details concerning this method. In general, this method is regarded as very efficient, however, it tends to create clusters of small size.

Ultrametric spaces provide a natural way to describe hierarchically structured complex systems, since the concept of ultrametricity is directly connected to the concept of hierarchy (Rammal et al (1986)).

Hierarchical trees associated with the single linkage between clusters can be obtained as follows. Let the distance matrix is given by table 1.

Table. 1. Distance matrix

|      | PG | IBM  | MSFT | INTC | KO   | JPM  |
|------|----|------|------|------|------|------|
| PG   | 0  | 1.15 | 1.18 | 1.15 | 0.47 | 0.64 |
| IBM  |    | 0    | 0.60 | 0.64 | 1.26 | 1.16 |
| MSFT |    |      | 0    | 0.45 | 1.27 | 1.11 |
| INTC |    |      |      | 0    | 1.26 | 0.74 |
| KO   |    |      |      |      | 0    | 0.94 |
| JPM  |    |      |      |      |      | 0    |

To obtain hierarchical tree we find the pair of stocks separated by the smallest distance: INTC and MSFT ($d$=0.45). Then find the pair of stocks with the next-smallest distance: PG and KO ($d$=0.47). Thus we have two separated clusters (KO-PG and INTC-MSFT). If we continue, we find MSFT and IBM ($d$=0.60). At this point our clusters are follows KO-PG and INTC-MSFT-IBM. The next pairs of closest stocks are JPM-PG and INTC-IBM ($d$=0.64). So the two clusters are INTC-MSFT-IBM and JPM-KO-PG. The smallest distance connecting the two clusters is observed for JPM-INTC ($d$=0.74). Table 2 gives matrix of the ultrametric distances.

Table. 2. Matrix of the ultrametric distances

|      | PG | IBM  | MSFT | INTC | KO   | JPM  |
|------|----|------|------|------|------|------|
| PG   | 0  | 0.74 | 0.74 | 0.74 | 0.47 | 0.64 |
| IBM  |    | 0    | 0.60 | 0.60 | 0.74 | 0.74 |
| MSFT |    |      | 0    | 0.45 | 0.74 | 0.74 |
| INTC |    |      |      | 0    | 0.74 | 0.74 |
| KO   |    |      |      |      | 0    | 0.64 |
| JPM  |    |      |      |      |      | 0    |

In Fig 5 we show the hierarchical tree obtained by the ultrametric method described above (see, also (Mantegna, Stanley (2001))).

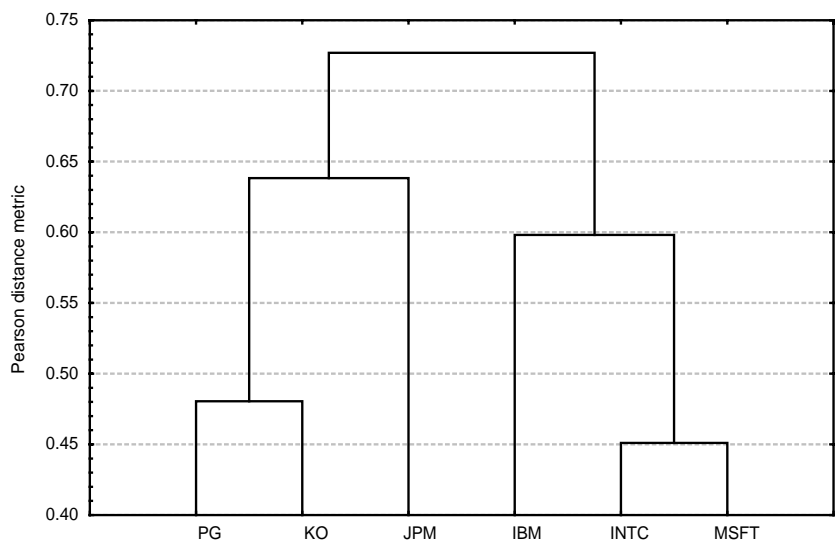

Figure 5.
Hierarchical tree.

However this method have some demerits. This caused by variety of distances and distance rules that affect on cluster structure of the hierarchical tree.

# 4 Study of Dow-Jones Industrial Average Index by ultrametric method

In this section we investigate DJIA by ultrametric method. We obtain cluster structure of this index by building appropriate hierarchical trees.

Let us first normalize $S_i(t)$ given by equation (2) on the interval [0,1] by nonlinear transformation $\tilde{S}_i = f(\frac{S_i - \bar{S}_i}{\sigma_i})$. Here $f(a) = \frac{1}{1+e^{-a}}$ is the activation function, $\bar{S}_i = \frac{1}{P}\sum_{j=1}^{P} S_i^j$ is an average of distribution and $\sigma_i = \frac{1}{P-1}\sum_{j=1}^{P}(S_i^j - \bar{S}_i)^2$ is a root-mean-square deviation (volatility) of the i-th stock.

In this study we use hourly data for DJIA. For the set of 30 stocks of the DJIA (table 3[1]), there are $(30 \times 29)/2 = 435$ different $\rho_{ij}$. All $\rho_{ij}$ are calculated for each studied time period. In Table 4 maximal and minimal values of $\rho_{ij}$ are summarized. Maximal value $\rho_{ij}$=0,61 from *10-Jan-94 to 27-Oct-97* corresponds to the companies JP Morgan Chase and American Express (Fig. 6). Maximal value $\rho_{ij}$=0,72 from *10-Nov-1997 to 27-Aug-2001* corresponds to the companies JP Morgan Chase and Citigroup (Fig.7). All this companies are engaged in financial services.

Table 3. The set of 30 stocks of the DJIA30.

| Ticker | Company | Industry sector |
|---|---|---|
| AA | Alcoa inc. | Metallurgy |
| AXP | American Express | Providing financial service |
| BA | Boeing | The aerospace industry |
| C | Citigroup | Providing financial service |
| CAT | Caterpillar | Mechanical engineering |
| DD | DuPont | Manufacture of hi-tech materials |
| DIS | Walt Disney | Showbiz industry |
| EK | Eastman Kodak | Imaging products and services |
| GE | General Electric | The electrotechnical company |
| GM | General Motors | Mechanical engineering |
| HD | Home Depot | Home improvement retailer |
| HON | Honeywell International | The aerospace industry |
| HWP | Hewlett-Packard | Computer technique industry |
| IBM | International Bus. Machine | Computer technique industry |
| INTC | Intel | Computer technique industry |
| IP | International Paper | Paper and packaging company |
| JNJ | Johnson & Johnson | Cosmetic industry |
| JPM | JP Morgan Chase | Providing financial service |
| KO | Coca Cola inc. | Foodstuff industry |
| MCD | McDonalds Corp. | Foodstuff industry |
| MMM | Minnesota Mining | Abrasives manufacturing |
| MO | Philip Morris | Tobacco industry |
| MRK | Merck & CO | Pharmaceutical products industry |
| MSFT | Microsoft | Software development |
| PG | Procter & Gamble | Cosmetic industry |
| SBC | SBC Communications | Providing of communication |
| T | AT&T | Providing of communication |
| UTX | United Technology | The aerospace industry |
| WMT | Wal-Mart Stores | Operation of mass merchandising stores |
| XOM | Exxon Mobil | Mining and sale of coal, copper and other minerals |

---

[1] finance.yahoo.com.

Table 4. The observed minimum and maximum values of correlation coefficient $\rho_{ij}$ for 30 stocks of DJIA.

| Time period | Min $\rho_{ij}$ | Max $\rho_{ij}$ |
|---|---|---|
| from 10-Jan-94 to 27-Oct-97 | -0,04 | 0,61 |
| from 10-Nov-97 to 27-Aug-01 | -0,06 | 0,72 |

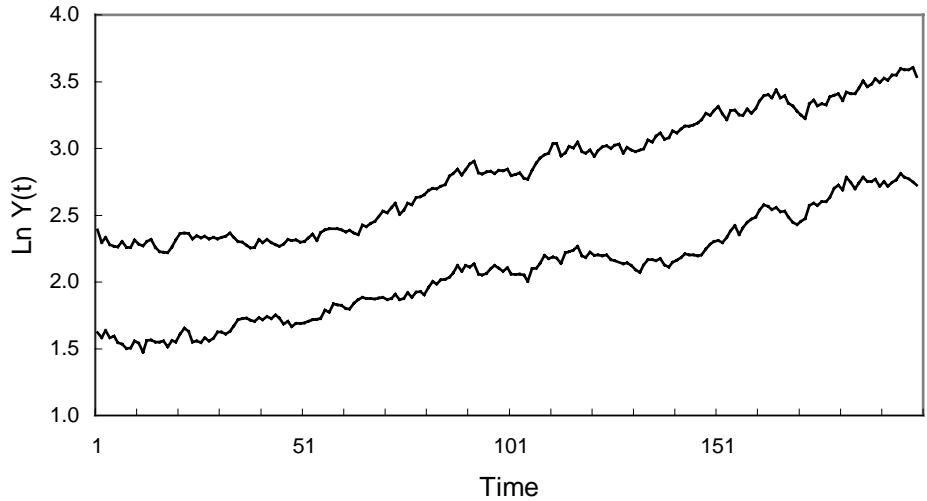

Figure 6.

Time evolution of ln Y(t) for JP Morgan Chase
(top curve) and American Express (bottom curve)
from *10-Jan-1994 to 27-Oct-1997*.

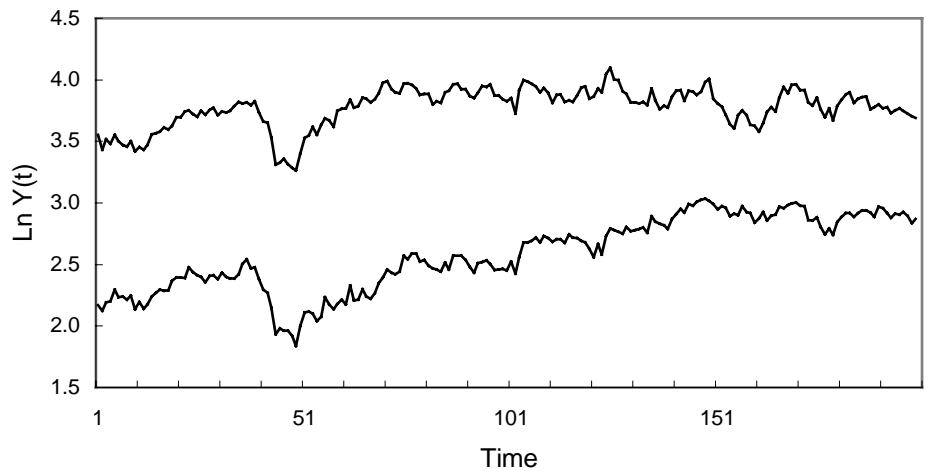

Figure 7.

Time evolution of ln Y(t) for JP Morgan Chase
(top curve) and Citigroup (bottom curve)
from *10-Nov-1997 to 27-Aug-2001*.

In Figures 8 and 9 hierarchical trees for 30 stocks which composed DJIA from *10-Jan-94 to 27-Oct-97* and from *10-Nov-97 to 27-Aug-01* respectively are represented. Hierarchical trees were constructed on the basis of Pearson distance metric and unweighted pair-group average linkage.

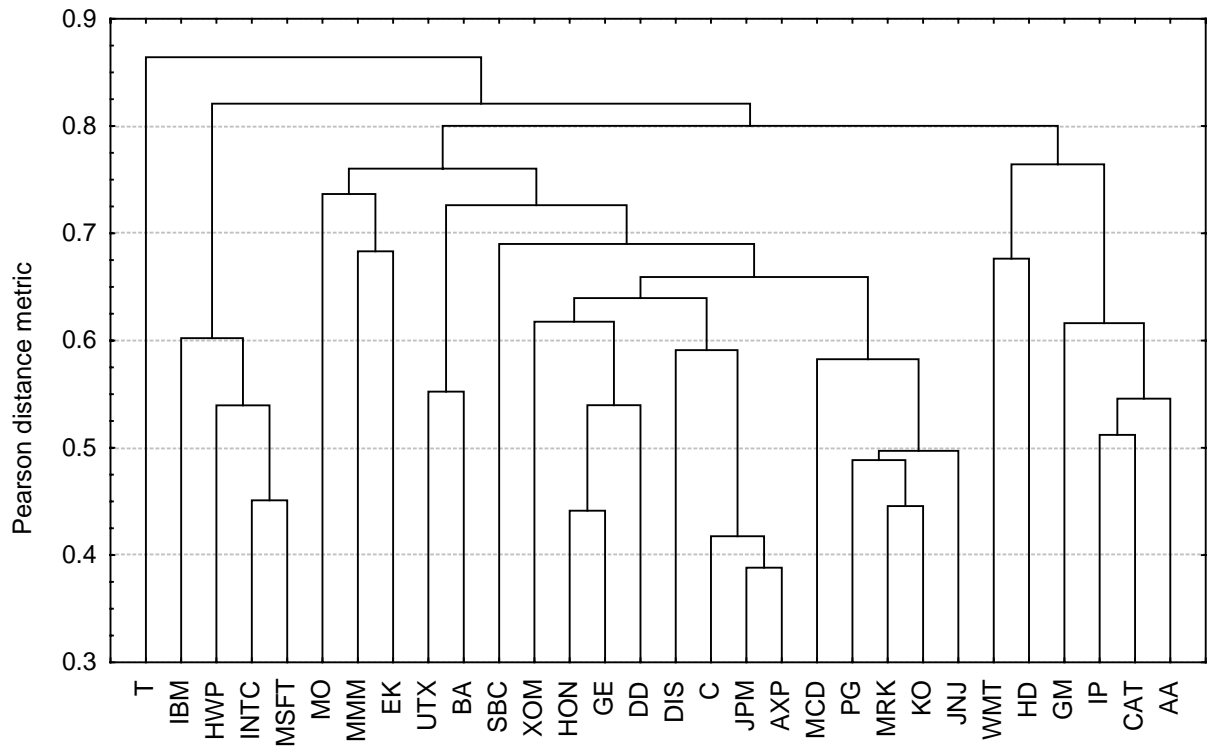

Figure 8.
Hierarchical tree for 30 stocks used to compute the DJIA
from *10-Jan-1994 to 27-Oct-1997*

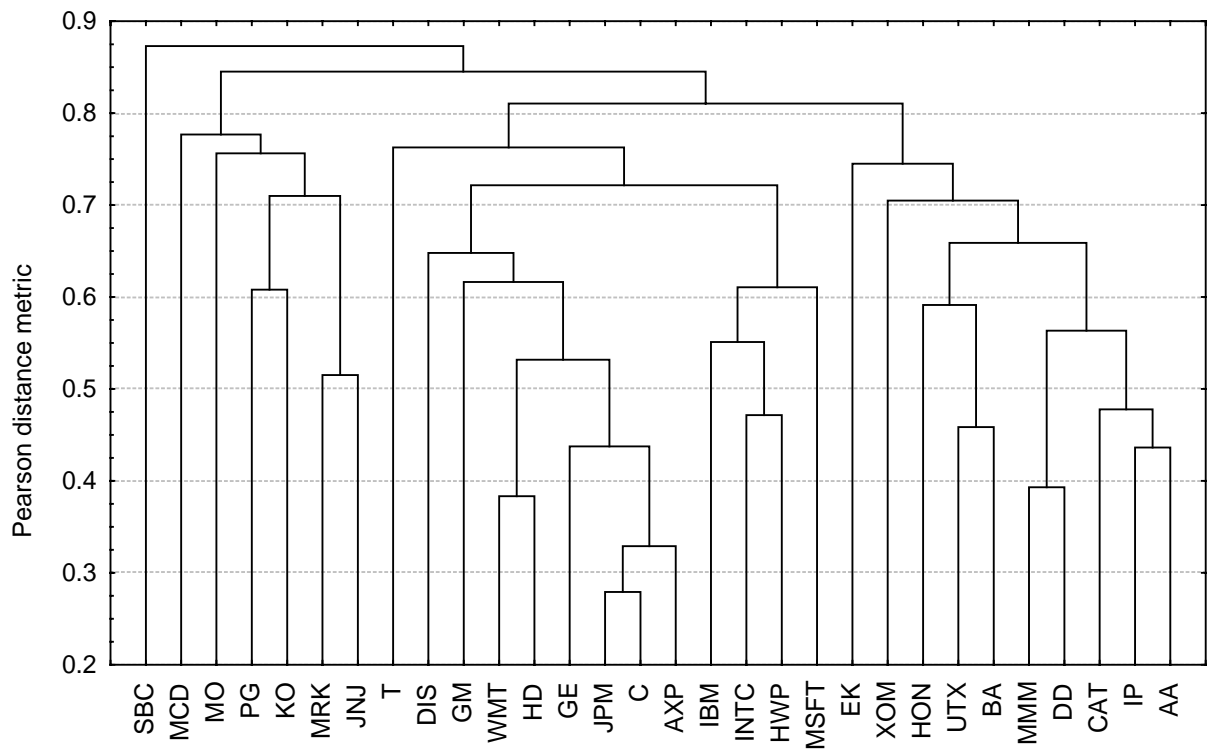

Figure 9.
Hierarchical tree for 30 stocks used to compute the DJIA
from *10-Nov-1997 to 27-Aug-2001*

Figures 8 and 9 represent cluster structure of the DJIA for each investigated period. As can be seen, financial companies AXP, C and JPM belong to the most correlated cluster. Although telecommunication companies T (Fig. 8.) and SBC (Fig. 9.) are weakly correlated with other companies.

# 5    Application of SOM method for clustering DJIA

In the present section we apply Self-Organized Maps method for clustering DJIA and we compare results of this method with results obtained by ultrametric method.
Construction of the SOM consists of two stages:

1) choice of parameters of training;
2) direct training of SOM.

Process of training includes rough tuning and exact fine-tuning. At a stage of rough tuning rough clusterization of the training set occurs. For this stage large correction of weights of the output layer is occurring. But at the stage of exact turning value of the correction considerably decreases. Experimentally we found that it is desirable, that the number of epochs for rough tuning was in 100 times more then neurons in a map.

The radius of training at the beginning of training should be commensurable with the size of a map, and at the end of training - small enough. Learning rate changes depending on the sizes of a map and the number of epochs. Learning rate can vary by the following rules:

o linear;
o inversely to the number of examples.

We used the second rule because it gives the smoothest change of the weights of neurons.

Initialization of weights of vectors of the SOM can be made by:

o random values;
o examples from the training set.

In the present study we used random values of initialization. Table 5 summaries the parameters of training of the SOM for DJIA.

Table 5. Parameters of training the SOM for DJIA from *10-Jan-94 to 27-Oct-97* and from *10-Nov-97 to 27-Aug-01*.

| The size of a map | 40x40 neurons |
|---|---|
| The form of cells | Hexagons |
| Number of epoch at the rough tuning | 100000 |
| Number of epoch at the exact tuning | 100000 |
| Learning rate at the rough tuning | 0.2 |
| Learning rate at the exact tuning | 0.05 |
| Initial radius of training | 20 |
| Final radius of training | 1 |
| Updating of the learning rate | Inversely proportional |
| Initial weights initialization | Random values |
| Time of training | 3 hours and 15 minutes |

Figures 10 and 11 show the learning error in log-linear coordinates. Namely, the logarithm of an average error of training (average distance between all weight factors of neurons in the SOM and actual values of training samples) versus the number of epochs is shown.

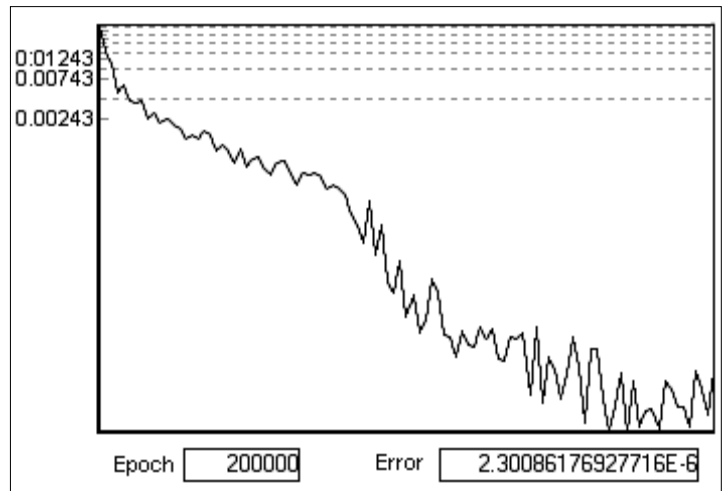

Figure 10.
Learning error for the SOM of DJIA
from *10-Jan-94 to 27-Oct-97*.

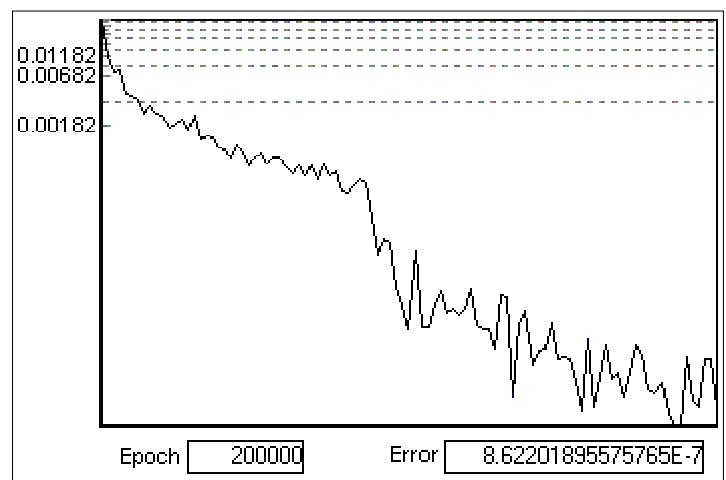

Figure 11.
Learning error for the SOM of DJIA
from *10-Nov-97 to 27-Aug-01*.

The U-matrix holds all distances between neighboring map units (Kaski (1997), Graepel et al (1998)). Dark blue color of a map unit means, that the average distance from this unit to its nearest neighbors is not enough. The darkest blue cells correspond to the neural-winners. Red color of a map unit means that the average distance from these cells to their nearest neighbors is large. Cluster is the group of vectors, distances between which inside this group is less, than distances to the next groups. The large difference between color of two clusters means, that these clusters are far apart. Figures 12 and 13 represent the U-matrix, cluster structure of SOM, and the error of clusterization for the both investigated periods. The Figure appropriate to the error of quantization of SOM, shows a distinction between modeling vectors of weights of neural-winners and actual input vectors. Red value of the map unit means, that the vector of weights of this unit is far from a real entrance vector.

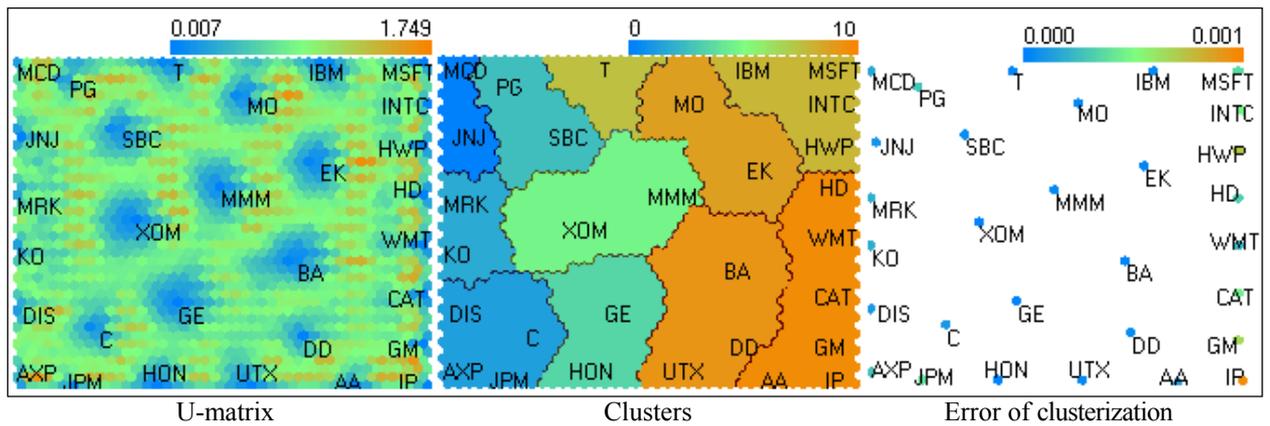

| U-matrix | Clusters | Error of clusterization |

Figure 12.
Self-organized map of DJIA index
from *10-Jan-94 to 27-Oct-97*.

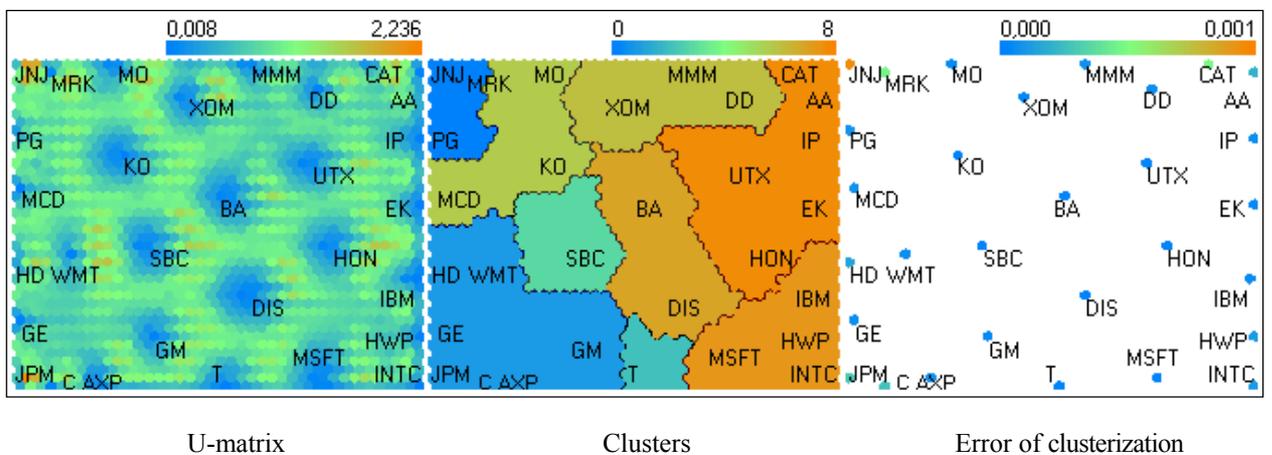

| U-matrix | Clusters | Error of clusterization |

Figure 13.
Self-organized map of DJIA index
from *10-Nov-97 to 27-Aug-01*.

The SOM divide all space of the companies on clusters, and everyone cluster includes the companies with high coefficient of correlation. Tables 6 – 8 summarize the correlation matrices for stocks, which are included into one cluster for the period from *10-Jan-94 to 27-Oct-97*. Appropriate clusters can be seen also in Figure 8.

Table 6. Correlation matrix for stocks which have got into one cluster from *10-Jan-94 to 27-Oct-97*

|     | AA | CAT | GM | HD | IP | WMT |
|-----|----|-----|-----|-----|-----|-----|
| AA  | 1  | 0,48 | 0,36 | 0,23 | 0,43 | 0,19 |
| CAT |    | 1   | 0,45 | 0,38 | 0,49 | 0,24 |
| GM  |    |     | 1   | 0,26 | 0,34 | 0,23 |
| HD  |    |     |     | 1   | 0,21 | 0,32 |
| IP  |    |     |     |     | 1   | 0,14 |
| WMT |    |     |     |     |     | 1   |

Table 7. Correlation matrix for stocks which have got into one cluster from *10-Jan-94 to 27-Oct-97*

|      | MSFT | HWP  | IBM  | INTC |
|------|------|------|------|------|
| MSFT | 1    | 0,40 | 0,40 | 0,55 |
| HWP  |      | 1    | 0,41 | 0,52 |
| IBM  |      |      | 1    | 0,38 |
| INTC |      |      |      | 1    |

Table 8. Correlation matrix for stocks which have got into one cluster from *10-Jan-94 to 27-Oct-97*

|     | AXP | C    | DIS  | JPM  |
|-----|-----|------|------|------|
| AXP | 1   | 0,57 | 0,40 | 0,61 |
| C   |     | 1    | 0,46 | 0,60 |
| DIS |     |      | 1    | 0,37 |
| JPM |     |      |      | 1    |

In Tables 9 - 11 correlation matrices for groups of stocks, which have fallen into three different clusters for the period from *10-Nov-97 to 27-Aug-01* are given.

Table 9. Correlation matrix for stocks which have fallen into one cluster for the period from *10-Nov-97 to 27-Aug-01*

|     | AA  | CAT  | EK   | HON  | IP   | UTX  |
|-----|-----|------|------|------|------|------|
| AA  | 1,  | 0,51 | 0,25 | 0,24 | 0,56 | 0,33 |
| CAT |     | 1    | 0,34 | 0,33 | 0,54 | 0,44 |
| EK  |     |      | 1    | 0,31 | 0,31 | 0,30 |
| HON |     |      |      | 1    | 0,41 | 0,47 |
| IP  |     |      |      |      | 1    | 0,36 |
| UTX |     |      |      |      |      | 1    |

Note, that shown in Figure 9 cluster BA-MMM-DD-AA-CAT-EK-HON-IP-UTX contains stocks from Table 9 and also stocks BA, MMM and DD from the nearest cluster of SOM.

Table 10. Correlation matrix for stocks which have fallen into one cluster from *10-Nov-97 to 27-Aug-01*.

|      | MSFT | HWP  | IBM  | INTC |
|------|------|------|------|------|
| MSFT | 1    | 0,38 | 0,31 | 0,48 |
| HWP  |      | 1    | 0,46 | 0,53 |
| IBM  |      |      | 1    | 0,44 |
| INTC |      |      |      | 1    |

Table 11. Correlation matrix for stocks which have fallen into one cluster from *10-Nov-97 to 27-Aug-01*.

|     | AXP | C    | GE   | GM   | HD   | WMT  | JPM  |
|-----|-----|------|------|------|------|------|------|
| AXP | 1   | 0,69 | 0,61 | 0,47 | 0,48 | 0,49 | 0,65 |
| C   |     | 1    | 0,57 | 0,45 | 0,49 | 0,52 | 0,72 |
| GE  |     |      | 1    | 0,38 | 0,49 | 0,51 | 0,51 |
| GM  |     |      |      | 1    | 0,33 | 0,25 | 0,42 |
| HD  |     |      |      |      | 1    | 0,62 | 0,42 |
| WMT |     |      |      |      |      | 1    | 0,36 |
| JPM |     |      |      |      |      |      | 1    |

The further two clusters from each other are (correspondingly the more difference in color between two clusters is), the less the correlation coefficient between prices of stocks within these clusters are. In Tables 12 and 13 correlation coefficients for the stocks, which have fallen into the darkest blue, and the reddest clusters are given for both studied periods.

Table 12. Correlation matrix for stocks which have fallen into two most distant clusters for the period from *10-Jan-94 to 27-Oct-97*.

|     | AA    | CAT  | GM   | HD   | IP   | WMT  |
|-----|-------|------|------|------|------|------|
| JNJ | -0,03 | 0,16 | 0,14 | 0,19 | 0,07 | 0,16 |
| MCD | 0,04  | 0,20 | 0,04 | 0,17 | 0,05 | 0,07 |

Table 13. Correlation matrix for stocks which have fallen into two most distant clusters for the period from *10-Nov-97 to 27-Aug-01*.

|     | AA    | CAT  | EK   | HON   | IP   | UTX  |
|-----|-------|------|------|-------|------|------|
| JNJ | 0,07  | 0,17 | 0,04 | -0,04 | 0,07 | 0,21 |
| MRK | -0,01 | 0,14 | 0,02 | 0,13  | 0,10 | 0,25 |
| PG  | 0,06  | 0,16 | 0,08 | 0,14  | 0,07 | 0,28 |

The given statistics shows, that the method of ultrametric spaces and the SOM method are in fair agreement for the DJIA portfolio.

## 6 The study of the NASDAQ100 Index by ultrametric method

In this section we investigate NASDAQ100 by ultrametric method. We obtain cluster structure of this index by building hierarchical trees.

For 100 stocks which form the NASDAQ100 (see table 14[2]), there are $(100 \times 99)/2 = 4950$ different correlation coefficients $\rho_{ij}$. All $\rho_{ij}$ are calculated for investigated time period from *14-Mar-01 to 31-Dec-01*. In the case of NASDAQ100 we used daily data. Maximal value of $\rho_{ij}$=0,93 characterizes two companies, developing microprocessors, namely, Novellus Systems, Inc. (NVLS) and KLA-Tencor Corporation (KLAC), (the graphs of the logarithm of stock prices of these companies are shown in Figure 14.

Table 14. The list of the stocks forming the NASDAQ100 index.

| Ticker | Company | Industry sector |
|--------|---------|-----------------|
| AAPL | Apple Computer, Inc. | Computer technique development |
| ABGX | Abgenix, Inc. | Biopharmaceutical industry |
| ADBE | Adobe Systems Inc. | Software development |
| ADCT | ADC Telecommunications | Telecommunication provider |
| ADLAC | Adelphia Communications | Communication industry |
| ADRX | Andrx Corporation | Biopharmaceutical industry |
| ALTR | Altera Corp | Software development |
| AMAT | Applied Materials, Inc. | Semiconductor systems development |
| AMCC | Applied Micro Circuits | Optic nets development |
| AMGN | Amgen, Inc. | Pharmaceutics industry |
| AMZN | Amazon.com, Inc. | IT |
| APOL | Apollo Group, Inc. | Higher education adults development |

---

[2] finance.yahoo.com.

| Ticker | Company | Industry |
|---|---|---|
| ATML | Atmel Corporation | IC products manufacturing |
| BBBY | Bed Bath & Beyond Inc. | Home improvement retailer |
| BEAS | BEA Systems, Inc. | Software development |
| BGEN | Biogen, Inc. | Biopharmaceutical industry |
| BMET | Biomet, Inc. | Pharmaceutics industry |
| BRCD | Brocade Communications | Storage area networks infrastructure production |
| BRCM | Broadcom Corporation | IT |
| CDWC | CDW Computer Centers, Inc. | High integrated silicon solutions provider |
| CEFT | Concord EFS, Inc. | Multi-brand computers development |
| CEPH | Cephalon, Inc. | Biopharmaceutical industry |
| CHIR | Chiron Corporation | Biopharmaceutical industry |
| CHKP | Check Point Software Tech | Software development |
| CHTR | Charter Communications | Cable systems operator |
| CIEN | CIENA Corporation | Optical network production |
| CMCSK | Comcast Corporation | Communication industry |
| CMVT | Comverse Technology, Inc. | Computer technique development |
| CNXT | Conexant Systems Inc. | Semiconductor systems development |
| COST | Costco Wholesale Corp. | Trading corporation |
| CPWR | Compuware Corporation | Software development |
| CSCO | Cisco Systems, Inc. | Telecommunication provider |
| CTAS | Cintas Corporation | Ancillary service provider |
| CTXS | Citrix Systems, Inc. | Software development |
| CYTC | CYTYC Corporation | Development of preparation system for medical applications |
| DELL | Dell Computer Corporation | Computer technique development |
| DISH | EchoStar Communications | Telecommunication provider |
| EBAY | eBay Inc. | IT |
| ERICY | LM Ericsson (ADR) | Telecommunication provider |
| ERTS | Electronic Arts Inc. | Software development |
| ESRX | Express Scripts, Inc. | Health care management |
| FISV | Fiserv, Inc. | Information management systems development |
| FLEX | Flextronics International | Telecommunication provider |
| GENZ | Genzyme General Division | Biotechnology development |
| GILD | Gilead Sciences, Inc. | Biopharmaceutical industry |
| GMST | Gemstar-TV Guide Intern-l | Computer technique development |
| HGSI | Human Genome Sciences | Therapeutic product development |
| ICOS | ICOS Corporation | Therapeutic product development |
| IDPH | IDEC Pharmaceuticals Corp | Biopharmaceutical industry |
| IDTI | Integrated Device Tech. | Telecommunication provider |
| IMCL | ImClone Systems, Inc. | Biopharmaceutical industry |
| IMNX | Immunex Corporation | Biopharmaceutical industry |
| INTC | Intel Corporation | Computer technique development |
| INTU | Intuit, Inc. | Software development |
| ITWO | I2 Technologies, Inc. | Software development |
| IVGN | Invitrogen Corporation | Life science market products development |
| JDSU | JDS Uniphase Corporation | Fiber optic components development |
| JNPR | Juniper Networks, Inc. | Software development |
| KLAC | KLA-Tencor Corporation | Semiconductor and related microelectronics development |
| LLTC | Linear Technology Corp. | Integrated circuits production |
| MCHP | Microchip Technology Inc. | Semiconductor development |
| MEDI | MedImmune, Inc. | Biotechnology |
| MERQ | Mercury Interactive Corp. | Management solutions performance |
| MLNM | Millennium Pharmaceutical | Biopharmaceutical industry |
| MOLX | Molex, Inc. | Fiber optic interconnection products manufacturer |
| MSFT | Microsoft Corporation | Software development |
| MXIM | Maxim Integrated Products | Semiconductor and related microelectronics development |
| NTAP | Network Appliance, Inc. | Computer technique development |
| NVDA | NVIDIA Corporation | Computer technique development |
| NVLS | Novellus Systems, Inc. | Semiconductor and related microelectronics development |
| NXTL | Nextel Communications | Telecommunication provider |
| ORCL | Oracle Corporation | Software development |
| PAYX | Paychex, Inc. | Financial service provider |
| PCAR | PACCAR Inc. | Distribution of high-quality commercial trucks |
| PDLI | Protein Design Labs, Inc. | Biotechnology |
| PMCS | PMC-Sierra, Inc. | Semiconductor and related microelectronics development |
| PSFT | PeopleSoft, Inc. | Software development |
| QCOM | QUALCOMM, Inc. | Software development |
| QLGC | QLogic Corporation | Semiconductor and related microelectronics development |
| RATL | Rational Software Corp. | Software development |
| RFMD | RF Micro Devices, Inc. | Integrated circuits development |
| SANM | Sanmina-SCI Corporation | Integrated electronic manufacturing service provider |
| SBUX | Starbucks Corporation | High-quality whole bean coffee roaster |
| SEBL | Siebel Systems, Inc. | Software development |
| SEPR | Sepracor Inc. | Biopharmaceutical industry |
| SNPS | Synopsys, Inc. | Software development |
| SPLS | Staples, Inc. | Office products superstore and distributor |
| SPOT | PanAmSat Corporation | Telecommunication provider |
| SSCC | Smurfit-Stone Container | Paperboard and paper-based packaging manufacturer |
| SUNW | Sun Microsystems, Inc. | Computer technique development |
| SYMC | Symantec Corporation | Network security solutions provider |
| TLAB | Tellabs, Inc. | Telecommunication provider |
| TMPW | TMP Worldwide Inc. | Advertising agency |
| USAI | USA Interactive | IT |
| VRSN | VeriSign, Inc. | Telecommunication provider |
| VRTS | Veritas Software Corp. | Software development |
| VTSS | Vitesse Semiconductor | High-performance integrated circuits development |
| WCOM | WorldCom Group | Telecommunication provider |
| XLNX | Xilinx, Incorporated | Software development |
| YHOO | Yahoo! Inc. | IT |

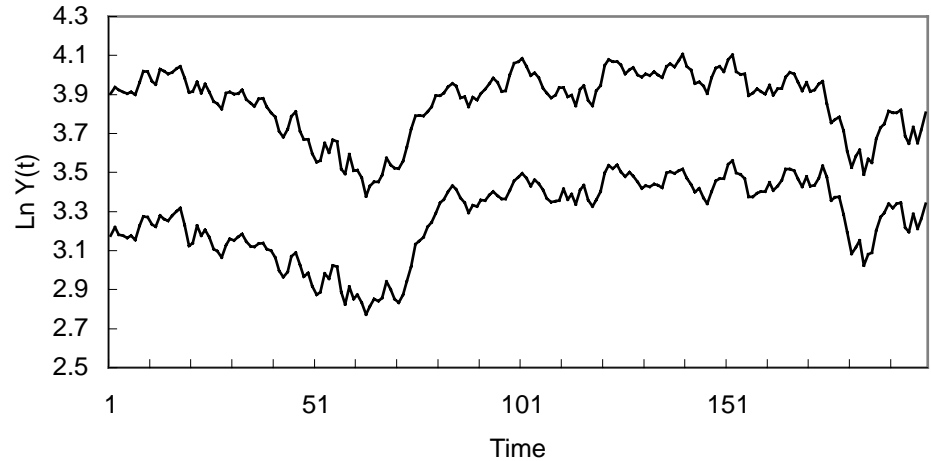

Figure 14.
Ln Y (t) of KLAC (the top graph) and NVLS (the bottom graph) companies
from *14-Mar-01 to 31-Dec-01*.

In Figure 15 the hierarchical tree for 100 companies form NASDAQ100 index is shown. Hierarchical trees were constructed by using the Pearson distance metric and unweighted pair-group average linkage.

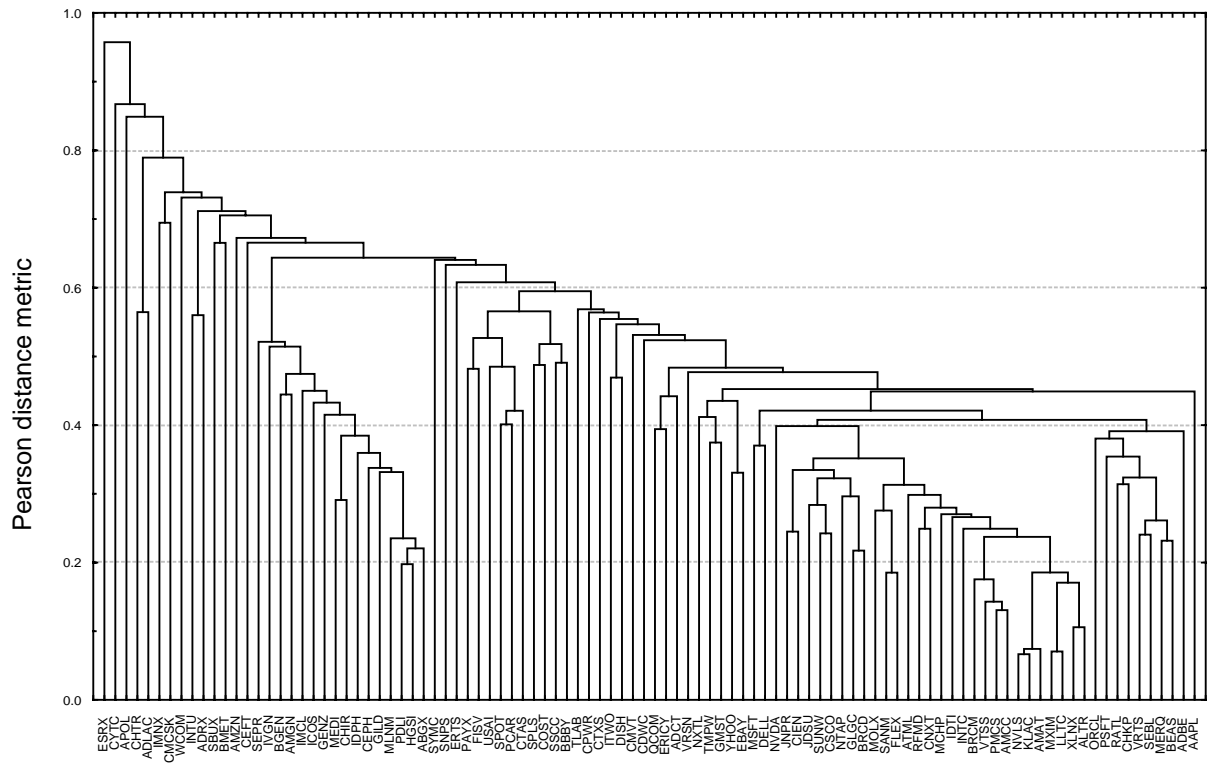

Figure 15.

The hierarchical tree for NASDAQ100 constructed by using the Pearson distance metric
and unweighted pair-group average connection between clusters.

In Figure 15 one can see 4 clusters:
- SERP-IVGN-BGEN-AMGN-IMCL-ICOS-GENZ-MEDI-CHIR-IDPH-CEPH-GILD-MLNM-PDLI-HGSI-ABG X;
- PAYX-FISV-USAI-SPOT-PCAR-CTAS-SPLS-COST-SSCC-BBBY;
- JNPR-CIEN-JDSU-SUNW-CSCO-NTAP-GLGC-BRCD-MOLX-SANM-FLEX-ATML-RFMD-CNXT-MCHP-IDT I-INTC-BRCM-VTSS-PMCS-AMCC-NVLS-KLAC-AM AT-MXIM-LLTC-XLNX-ALTR
- ORCL-PSFT-RATL-CHKP-VRTS-SEBL-MERQ-BEAS-ADBE.

The most correlated cluster KLAC-NVLS-AMAT consists of the companies developing microprocessors.

## 7  Application of SOM method for clustering NASDAQ100

In this section we present the results obtained by neural network technology for clustering NASDAQ100 portfolio. Since the details of the approach have been described in previous sections here we restrict ourselves by expositions of results only. As quantization of NASDAQ100 is quite similar to DJIA we omit all comments and discussion, which already have been done for DJIA. In Table 15 the training parameters of SOM for NASDAQ100 are given.

Table 15. Training parameters of SOM for NASDAQ100 during the time period *14-Mar-01 to 31-Dec-01*

| | |
|---|---|
| The size of a map | 40x40 neurons |
| The form of cells | Hexagons |
| Number of epoch at the rough tuning | 100000 |
| Number of epoch at the exact tuning | 100000 |
| Learning rate at the rough tuning | 0.3 |
| Learning rate at the exact tuning | 0.05 |
| Initial radius of training | 20 |
| Final radius of training | 1 |
| Updating of the learning rate | Inversely proportional |
| Initial weights initialization | Random values |
| Time of training | 4 hours and 10 minutes |

In Figures 16-19 the learning error, cluster structure of SOM, U-matrix, and the error of clusterization are shown.

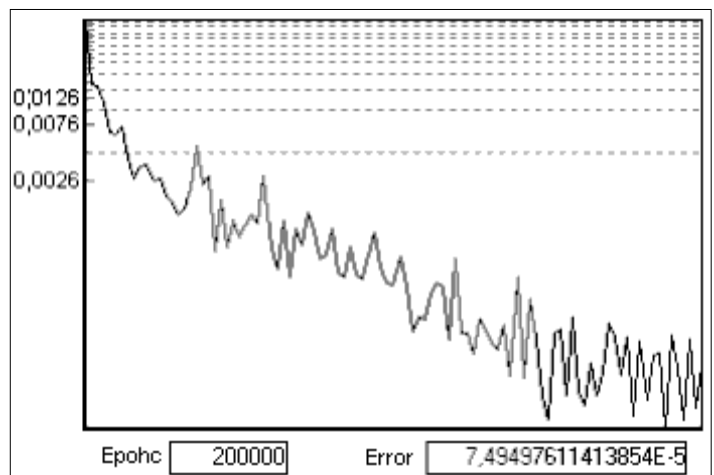

Figure 16

Learning error of SOM for NASDAQ100
during the period 10-Nov-97 to 27-Aug-01.

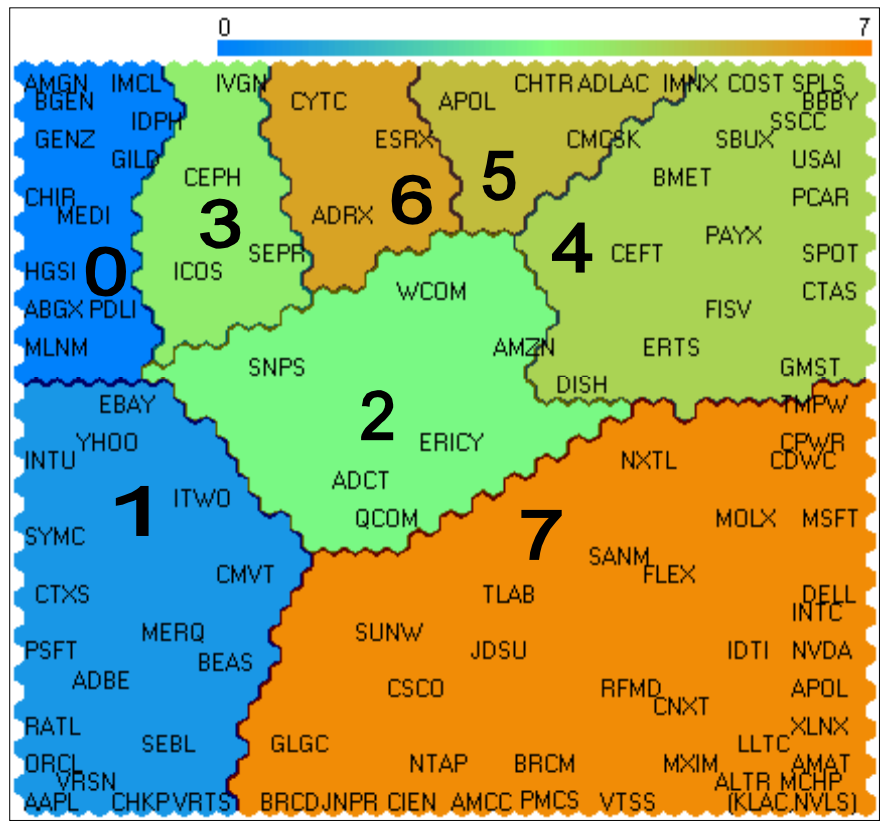

Figure 17.
Cluster structure of SOM.

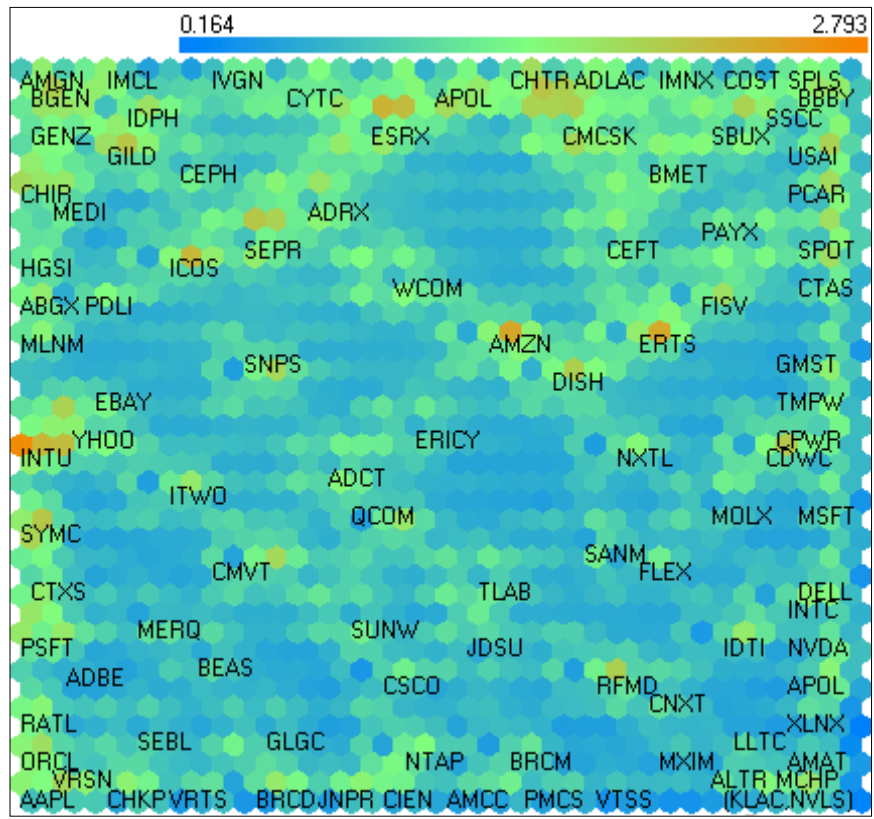

Figure 18. U-matrix.

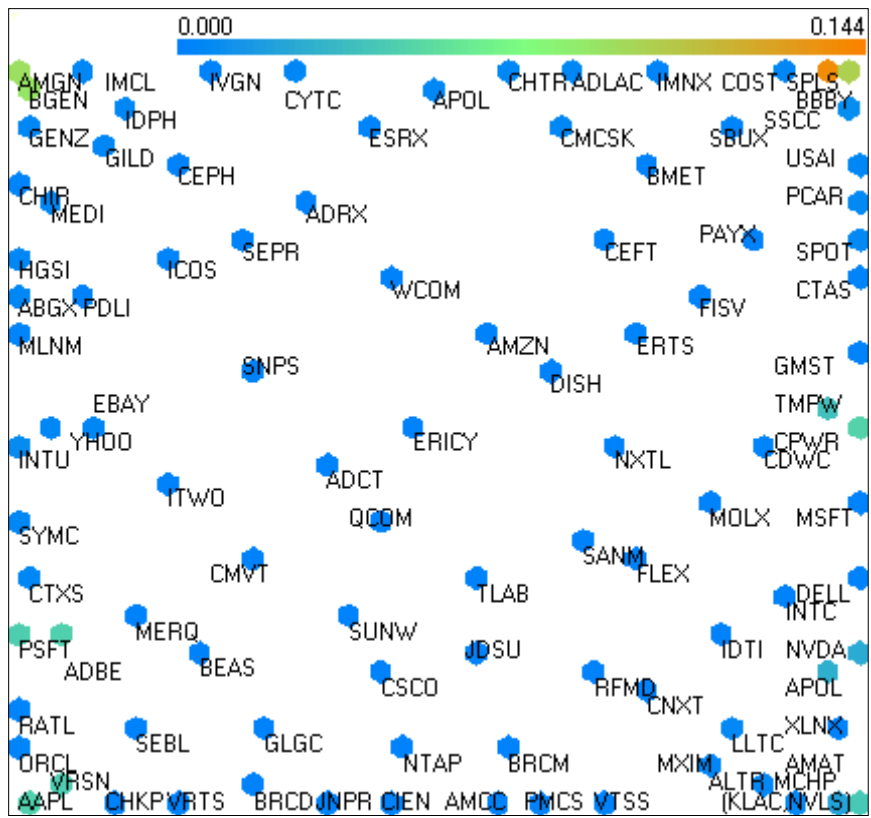

Figure 19. Error of clusterization.

One can see that due to the insufficient number of neurons (cells of the map) the U-matrix has a fuzzy structure (Fig.18). For the same reason, the more correlated stocks KLAC and NVLS have fallen into one cell. However, the statistics given below (Tables16-22), proves the successfulness of clusterization.

Table 16. The correlation matrix for stocks fallen into the cluster № 0 (see Fig.17).

|      | ABGX | AMGN | BGEN | CHIR | GENZ | GILD | HGSI | IDPH | IMCL | MEDI | MLNM | PDLI |
|------|------|------|------|------|------|------|------|------|------|------|------|------|
| ABGX | 1    | 0.51 | 0.50 | 0.61 | 0.57 | 0.68 | 0.77 | 0.66 | 0.59 | 0.67 | 0.75 | 0.78 |
| AMGN |      | 1    | 0.56 | 0.56 | 0.49 | 0.54 | 0.59 | 0.57 | 0.43 | 0.61 | 0.53 | 0.54 |
| BGEN |      |      | 1    | 0.55 | 0.56 | 0.50 | 0.61 | 0.46 | 0.44 | 0.62 | 0.54 | 0.54 |
| CHIR |      |      |      | 1    | 0.60 | 0.59 | 0.66 | 0.57 | 0.46 | 0.71 | 0.55 | 0.61 |
| GENZ |      |      |      |      | 1    | 0.62 | 0.64 | 0.53 | 0.49 | 0.57 | 0.57 | 0.61 |
| GILD |      |      |      |      |      | 1    | 0.67 | 0.64 | 0.59 | 0.62 | 0.64 | 0.68 |
| HGSI |      |      |      |      |      |      | 1    | 0.67 | 0.59 | 0.69 | 0.79 | 0.80 |
| IDPH |      |      |      |      |      |      |      | 1    | 0.58 | 0.59 | 0.59 | 0.69 |
| IMCL |      |      |      |      |      |      |      |      | 1    | 0.51 | 0.58 | 0.63 |
| MEDI |      |      |      |      |      |      |      |      |      | 1    | 0.70 | 0.66 |
| MLNM |      |      |      |      |      |      |      |      |      |      | 1    | 0.75 |
| PDLI |      |      |      |      |      |      |      |      |      |      |      | 1    |

Table 17. Correlation matrix for stocks fallen into the cluster № 1.

|      | AAPL | ADBE | BEAS | CHKP | CMVT | CTXS | EBAY | INTU | ITWO | MERQ | ORCL | PSFT | RATL | SEBL | SYMC | VRSN | VRTS | YHOO |
|------|------|------|------|------|------|------|------|------|------|------|------|------|------|------|------|------|------|------|
| AAPL | 1    | 0.45 | 0.32 | 0.50 | 0.37 | 0.43 | 0.48 | 0.44 | 0.46 | 0.56 | 0.57 | 0.54 | 0.56 | 0.60 | 0.35 | 0.50 | 0.61 | 0.55 |
| ADBE |      | 1    | 0.54 | 0.44 | 0.32 | 0.54 | 0.58 | 0.61 | 0.43 | 0.41 | 0.36 | 0.45 | 0.41 | 0.35 | 0.52 | 0.45 | 0.41 | 0.34 |
| BEAS |      |      | 1    | 0.52 | 0.64 | 0.23 | 0.53 | 0.63 | 0.51 | 0.64 | 0.42 | 0.62 | 0.52 | 0.56 | 0.53 | 0.42 | 0.65 | 0.58 |
| CHKP |      |      |      | 1    | 0.49 | 0.48 | 0.56 | 0.38 | 0.49 | 0.71 | 0.58 | 0.58 | 0.69 | 0.69 | 0.44 | 0.66 | 0.68 | 0.59 |
| CMVT |      |      |      |      | 1    | 0.38 | 0.42 | 0.28 | 0.41 | 0.61 | 0.43 | 0.40 | 0.43 | 0.45 | 0.21 | 0.46 | 0.47 | 0.49 |
| CTXS |      |      |      |      |      | 1    | 0.47 | 0.34 | 0.45 | 0.59 | 0.44 | 0.49 | 0.52 | 0.49 | 0.28 | 0.42 | 0.53 | 0.52 |
| EBAY |      |      |      |      |      |      | 1    | 0.39 | 0.50 | 0.63 | 0.49 | 0.46 | 0.60 | 0.58 | 0.38 | 0.51 | 0.53 | 0.67 |
| INTU |      |      |      |      |      |      |      | 1    | 0.40 | 0.46 | 0.46 | 0.50 | 0.46 | 0.47 | 0.37 | 0.40 | 0.44 | 0.41 |
| ITWO |      |      |      |      |      |      |      |      | 1    | 0.59 | 0.46 | 0.56 | 0.57 | 0.59 | 0.41 | 0.47 | 0.57 | 0.56 |
| MERQ |      |      |      |      |      |      |      |      |      | 1    | 0.64 | 0.70 | 0.73 | 0.76 | 0.53 | 0.55 | 0.73 | 0.66 |
| ORCL |      |      |      |      |      |      |      |      |      |      | 1    | 0.54 | 0.64 | 0.68 | 0.40 | 0.56 | 0.61 | 0.58 |
| PSFT |      |      |      |      |      |      |      |      |      |      |      | 1    | 0.61 | 0.71 | 0.52 | 0.53 | 0.64 | 0.53 |
| RATL |      |      |      |      |      |      |      |      |      |      |      |      | 1    | 0.67 | 0.39 | 0.58 | 0.61 | 0.61 |
| SEBL |      |      |      |      |      |      |      |      |      |      |      |      |      | 1    | 0.47 | 0.63 | 0.76 | 0.62 |
| SYMC |      |      |      |      |      |      |      |      |      |      |      |      |      |      | 1    | 0.35 | 0.41 | 0.41 |
| VRSN |      |      |      |      |      |      |      |      |      |      |      |      |      |      |      | 1    | 0.64 | 0.60 |
| VRTS |      |      |      |      |      |      |      |      |      |      |      |      |      |      |      |      | 1    | 0.62 |
| YHOO |      |      |      |      |      |      |      |      |      |      |      |      |      |      |      |      |      | 1    |

Table 17 includes both all stocks fallen into the most left cluster of the hierarchical tree (Fig. 15), and companies INTU, SYMC, CTXS, ITWO, CMVT, YAHOO, EBAY which are scattered on the entire hierarchical tree. It shows the interesting effect, that on the large data sets the results of linear methods (in particular, a method of ultrametric spaces) are in essential disagreement with results of nonlinear methods of SOM.

Table 18. Correlation matrix for stocks fallen into the cluster № 2.

|       | ADCT | AMZN | ERICY | QCOM | SNPS | WCOM |
|-------|------|------|-------|------|------|------|
| ADCT  | 1    | 0.33 | 0.51  | 0.61 | 0.34 | 0.25 |
| AMZN  |      | 1    | 0.35  | 0.32 | 0.23 | 0.21 |
| ERICY |      |      | 1     | 0.61 | 0.42 | 0.39 |
| QCOM  |      |      |       | 1    | 0.39 | 0.36 |
| SNPS  |      |      |       |      | 1    | 0.22 |
| WCOM  |      |      |       |      |      | 1    |

Table 19. Correlation matrix for stocks fallen into the cluster № 3.

|      | CEPH | ICOS | IVGN | SEPR |
|------|------|------|------|------|
| CEPH | 1    | 0.62 | 0.55 | 0.51 |
| ICOS |      | 1    | 0.51 | 0.48 |
| IVGN |      |      | 1    | 0.44 |
| SEPR |      |      |      | 1    |

Table 20. Correlation matrix for stocks fallen into the cluster № 4.

|      | BBBY | BMET | CEFT | COST | CTAS | DISH | ERTS | FISV | GMST | PAYX | PCAR | SBUX | SPLS | SPOT | SSCC | USAI |
|------|------|------|------|------|------|------|------|------|------|------|------|------|------|------|------|------|
| BBBY | 1    | 0.39 | 0.25 | 0.50 | 0.48 | 0.39 | 0.34 | 0.43 | 0.56 | 0.45 | 0.54 | 0.39 | 0.49 | 0.37 | 0.51 | 0.44 |
| BMET |      | 1    | 0.32 | 0.29 | 0.38 | 0.27 | 0.31 | 0.39 | 0.29 | 0.31 | 0.39 | 0.33 | 0.29 | 0.44 | 0.31 | 0.36 |
| CEFT |      |      | 1    | 0.30 | 0.36 | 0.31 | 0.33 | 0.46 | 0.29 | 0.38 | 0.43 | 0.32 | 0.30 | 0.37 | 0.26 | 0.27 |
| COST |      |      |      | 1    | 0.51 | 0.33 | 0.21 | 0.33 | 0.48 | 0.44 | 0.48 | 0.37 | 0.51 | 0.43 | 0.44 | 0.43 |
| CTAS |      |      |      |      | 1    | 0.39 | 0.41 | 0.45 | 0.53 | 0.49 | 0.58 | 0.44 | 0.43 | 0.58 | 0.49 | 0.48 |
| DISH |      |      |      |      |      | 1    | 0.38 | 0.41 | 0.44 | 0.36 | 0.43 | 0.34 | 0.37 | 0.36 | 0.42 | 0.42 |
| ERTS |      |      |      |      |      |      | 1    | 0.42 | 0.41 | 0.34 | 0.35 | 0.20 | 0.25 | 0.41 | 0.28 | 0.33 |
| FISV |      |      |      |      |      |      |      | 1    | 0.45 | 0.52 | 0.57 | 0.43 | 0.28 | 0.53 | 0.36 | 0.37 |
| GMST |      |      |      |      |      |      |      |      | 1    | 0.38 | 0.54 | 0.39 | 0.45 | 0.50 | 0.50 | 0.55 |
| PAYX |      |      |      |      |      |      |      |      |      | 1    | 0.48 | 0.42 | 0.37 | 0.48 | 0.41 | 0.42 |
| PCAR |      |      |      |      |      |      |      |      |      |      | 1    | 0.46 | 0.50 | 0.60 | 0.55 | 0.53 |
| SBUX |      |      |      |      |      |      |      |      |      |      |      | 1    | 0.39 | 0.43 | 0.32 | 0.37 |
| SPLS |      |      |      |      |      |      |      |      |      |      |      |      | 1    | 0.34 | 0.49 | 0.45 |
| SPOT |      |      |      |      |      |      |      |      |      |      |      |      |      | 1    | 0.45 | 0.54 |
| SSCC |      |      |      |      |      |      |      |      |      |      |      |      |      |      | 1    | 0.47 |
| USAI |      |      |      |      |      |      |      |      |      |      |      |      |      |      |      | 1    |

Table 21. Correlation matrix for stocks fallen into the cluster № 5.

|       | ADLAC | APOL | CHTR | CMCSK | IMNX |
|-------|-------|------|------|-------|------|
| ADLAC | 1     | 0.13 | 0.44 | 0.32  | 0.22 |
| APOL  |       | 1    | 0.60 | 0.38  | 0.11 |
| CHTR  |       |      | 1    | 0.28  | 0.14 |
| CMCSK |       |      |      | 1     | 0.31 |
| IMNX  |       |      |      |       | 1    |

Table 22. Correlation matrix for stocks fallen into the cluster № 6.

|      | ADRX | CYTC | ESRX |
|------|------|------|------|
| ADRX | 1    | 0.31 | 0.30 |
| CYTC |      | 1    | 0.23 |
| ESRX |      |      | 1    |

Table 23. Correlation matrix for the stocks fallen into the cluster № 7

| | ALTR | AMAT | AMCC | APOL | BRCD | BRCM | CDWC | CIEN | CNXT | CPWR | CSCO | DELL | FLEX | IDTI | INTC | JDSU | JNPR | KLAC | LLTC | MCHP | MOLX | MSFT | MXIM | NTAP | NVDA | NVLS | NXTL | PMCS | GLGC | RFMD | SANM | SUNW | TLAB | TMPW | VTSS | XLNX |
|---|---|---|---|---|---|---|---|---|---|---|---|---|---|---|---|---|---|---|---|---|---|---|---|---|---|---|---|---|---|---|---|---|---|---|---|---|
| ALTR | 1 | 0.82 | 0.78 | 0.19 | 0.62 | 0.76 | 0.50 | 0.60 | 0.74 | 0.53 | 0.70 | 0.64 | 0.72 | 0.74 | 0.74 | 0.66 | 0.64 | 0.81 | 0.83 | 0.80 | 0.74 | 0.60 | 0.82 | 0.65 | 0.64 | 0.82 | 0.61 | 0.42 | 0.75 | 0.72 | 0.70 | 0.64 | 0.53 | 0.58 | 0.78 | 0.89 |
| AMAT | 0.82 | 1 | 0.77 | 0.18 | 0.58 | 0.74 | 0.48 | 0.58 | 0.73 | 0.48 | 0.67 | 0.66 | 0.73 | 0.74 | 0.80 | 0.65 | 0.60 | 0.93 | 0.84 | 0.73 | 0.65 | 0.59 | 0.81 | 0.68 | 0.66 | 0.92 | 0.57 | 0.45 | 0.76 | 0.67 | 0.68 | 0.66 | 0.49 | 0.52 | 0.74 | 0.85 |
| AMCC | 0.78 | 0.77 | 1 | 0.15 | 0.65 | 0.82 | 0.45 | 0.72 | 0.73 | 0.46 | 0.75 | 0.61 | 0.68 | 0.73 | 0.74 | 0.76 | 0.75 | 0.76 | 0.77 | 0.66 | 0.67 | 0.61 | 0.77 | 0.72 | 0.64 | 0.73 | 0.59 | 0.46 | 0.77 | 0.73 | 0.68 | 0.65 | 0.53 | 0.58 | 0.87 | 0.78 |
| APOL | 0.19 | 0.18 | 0.15 | 1 | 0.17 | 0.13 | 0.15 | 0.18 | 0.19 | 0.15 | 0.21 | 0.18 | 0.28 | 0.15 | 0.19 | 0.13 | 0.20 | 0.19 | 0.17 | 0.21 | 0.25 | 0.14 | 0.15 | 0.16 | 0.24 | 0.21 | 0.25 | 0.16 | 0.20 | 0.24 | 0.21 | 0.21 | 0.09 | 0.17 | 0.15 | 0.20 |
| BRCD | 0.62 | 0.58 | 0.65 | 0.17 | 1 | 0.71 | 0.45 | 0.64 | 0.58 | 0.40 | 0.70 | 0.54 | 0.61 | 0.64 | 0.56 | 0.62 | 0.69 | 0.60 | 0.67 | 0.56 | 0.55 | 0.52 | 0.61 | 0.70 | 0.55 | 0.56 | 0.55 | 0.31 | 0.78 | 0.59 | 0.55 | 0.66 | 0.41 | 0.47 | 0.64 | 0.63 |
| BRCM | 0.76 | 0.74 | 0.82 | 0.13 | 0.71 | 1 | 0.47 | 0.69 | 0.74 | 0.43 | 0.75 | 0.61 | 0.70 | 0.73 | 0.73 | 0.74 | 0.73 | 0.73 | 0.76 | 0.70 | 0.68 | 0.61 | 0.76 | 0.72 | 0.63 | 0.73 | 0.65 | 0.41 | 0.77 | 0.72 | 0.71 | 0.67 | 0.49 | 0.58 | 0.82 | 0.74 |
| CDWC | 0.50 | 0.48 | 0.45 | 0.15 | 0.45 | 0.47 | 1 | 0.34 | 0.50 | 0.43 | 0.47 | 0.54 | 0.59 | 0.50 | 0.61 | 0.46 | 0.44 | 0.43 | 0.50 | 0.45 | 0.62 | 0.53 | 0.46 | 0.46 | 0.38 | 0.42 | 0.43 | 0.39 | 0.46 | 0.43 | 0.52 | 0.52 | 0.34 | 0.52 | 0.47 | 0.50 |
| CIEN | 0.60 | 0.58 | 0.72 | 0.18 | 0.64 | 0.69 | 0.34 | 1 | 0.56 | 0.35 | 0.67 | 0.41 | 0.56 | 0.55 | 0.55 | 0.71 | 0.75 | 0.60 | 0.56 | 0.50 | 0.57 | 0.48 | 0.58 | 0.65 | 0.50 | 0.56 | 0.46 | 0.32 | 0.65 | 0.62 | 0.59 | 0.58 | 0.52 | 0.45 | 0.69 | 0.61 |
| CNXT | 0.74 | 0.73 | 0.73 | 0.19 | 0.58 | 0.74 | 0.50 | 0.56 | 1 | 0.46 | 0.66 | 0.56 | 0.70 | 0.72 | 0.71 | 0.66 | 0.61 | 0.74 | 0.75 | 0.70 | 0.68 | 0.53 | 0.74 | 0.64 | 0.57 | 0.75 | 0.61 | 0.46 | 0.68 | 0.75 | 0.65 | 0.67 | 0.45 | 0.53 | 0.75 | 0.75 |
| CPWR | 0.53 | 0.48 | 0.46 | 0.15 | 0.40 | 0.43 | 0.43 | 0.35 | 0.46 | 1 | 0.50 | 0.43 | 0.48 | 0.49 | 0.53 | 0.42 | 0.39 | 0.47 | 0.49 | 0.50 | 0.51 | 0.47 | 0.49 | 0.42 | 0.44 | 0.49 | 0.52 | 0.35 | 0.47 | 0.37 | 0.44 | 0.44 | 0.37 | 0.50 | 0.44 | 0.44 |
| CSCO | 0.70 | 0.67 | 0.75 | 0.21 | 0.70 | 0.75 | 0.47 | 0.67 | 0.66 | 0.50 | 1 | 0.61 | 0.66 | 0.67 | 0.69 | 0.76 | 0.70 | 0.63 | 0.70 | 0.56 | 0.65 | 0.55 | 0.71 | 0.70 | 0.54 | 0.59 | 0.57 | 0.44 | 0.72 | 0.64 | 0.64 | 0.76 | 0.55 | 0.54 | 0.72 | 0.68 |
| DELL | 0.64 | 0.66 | 0.61 | 0.18 | 0.54 | 0.61 | 0.54 | 0.41 | 0.56 | 0.43 | 0.61 | 1 | 0.59 | 0.62 | 0.71 | 0.58 | 0.53 | 0.62 | 0.70 | 0.58 | 0.64 | 0.63 | 0.66 | 0.55 | 0.61 | 0.61 | 0.52 | 0.43 | 0.62 | 0.52 | 0.53 | 0.56 | 0.39 | 0.57 | 0.57 | 0.66 |
| FLEX | 0.72 | 0.73 | 0.68 | 0.28 | 0.61 | 0.70 | 0.59 | 0.56 | 0.70 | 0.48 | 0.66 | 0.59 | 1 | 0.69 | 0.74 | 0.67 | 0.62 | 0.69 | 0.70 | 0.69 | 0.72 | 0.59 | 0.68 | 0.64 | 0.53 | 0.73 | 0.68 | 0.50 | 0.71 | 0.68 | 0.81 | 0.67 | 0.49 | 0.62 | 0.71 | 0.71 |
| IDTI | 0.74 | 0.74 | 0.73 | 0.15 | 0.64 | 0.73 | 0.50 | 0.55 | 0.72 | 0.49 | 0.67 | 0.62 | 0.69 | 1 | 0.70 | 0.63 | 0.62 | 0.73 | 0.80 | 0.72 | 0.62 | 0.53 | 0.76 | 0.66 | 0.64 | 0.72 | 0.65 | 0.46 | 0.74 | 0.67 | 0.65 | 0.67 | 0.49 | 0.56 | 0.74 | 0.73 |
| INTC | 0.74 | 0.80 | 0.74 | 0.19 | 0.56 | 0.73 | 0.61 | 0.55 | 0.71 | 0.53 | 0.69 | 0.71 | 0.74 | 0.70 | 1 | 0.68 | 0.58 | 0.74 | 0.79 | 0.69 | 0.74 | 0.69 | 0.75 | 0.63 | 0.64 | 0.74 | 0.59 | 0.41 | 0.69 | 0.66 | 0.69 | 0.70 | 0.45 | 0.59 | 0.72 | 0.77 |
| JDSU | 0.66 | 0.65 | 0.76 | 0.13 | 0.62 | 0.74 | 0.46 | 0.71 | 0.66 | 0.42 | 0.76 | 0.58 | 0.67 | 0.63 | 0.68 | 1 | 0.67 | 0.61 | 0.68 | 0.58 | 0.65 | 0.52 | 0.71 | 0.67 | 0.53 | 0.59 | 0.55 | 0.43 | 0.68 | 0.64 | 0.68 | 0.68 | 0.60 | 0.54 | 0.73 | 0.66 |
| JNPR | 0.64 | 0.60 | 0.75 | 0.20 | 0.69 | 0.73 | 0.44 | 0.75 | 0.61 | 0.39 | 0.70 | 0.53 | 0.62 | 0.62 | 0.58 | 0.67 | 1 | 0.61 | 0.64 | 0.53 | 0.61 | 0.52 | 0.66 | 0.65 | 0.55 | 0.58 | 0.57 | 0.38 | 0.74 | 0.64 | 0.62 | 0.62 | 0.48 | 0.54 | 0.70 | 0.60 |
| KLAC | 0.81 | 0.93 | 0.76 | 0.19 | 0.60 | 0.73 | 0.43 | 0.60 | 0.74 | 0.47 | 0.63 | 0.62 | 0.69 | 0.73 | 0.74 | 0.61 | 0.61 | 1 | 0.82 | 0.76 | 0.63 | 0.55 | 0.78 | 0.66 | 0.63 | 0.93 | 0.57 | 0.42 | 0.75 | 0.70 | 0.65 | 0.62 | 0.49 | 0.50 | 0.76 | 0.83 |
| LLTC | 0.83 | 0.84 | 0.77 | 0.17 | 0.67 | 0.76 | 0.50 | 0.56 | 0.75 | 0.49 | 0.70 | 0.70 | 0.70 | 0.80 | 0.79 | 0.68 | 0.64 | 0.82 | 1 | 0.74 | 0.70 | 0.61 | 0.93 | 0.66 | 0.72 | 0.80 | 0.61 | 0.45 | 0.81 | 0.73 | 0.70 | 0.66 | 0.48 | 0.59 | 0.78 | 0.84 |
| MCHP | 0.80 | 0.73 | 0.66 | 0.21 | 0.56 | 0.70 | 0.45 | 0.50 | 0.70 | 0.50 | 0.56 | 0.58 | 0.69 | 0.72 | 0.69 | 0.58 | 0.53 | 0.76 | 0.74 | 1 | 0.67 | 0.50 | 0.71 | 0.55 | 0.58 | 0.80 | 0.61 | 0.42 | 0.66 | 0.67 | 0.61 | 0.55 | 0.49 | 0.48 | 0.72 | 0.76 |
| MOLX | 0.74 | 0.65 | 0.67 | 0.25 | 0.55 | 0.68 | 0.62 | 0.57 | 0.68 | 0.51 | 0.65 | 0.64 | 0.72 | 0.62 | 0.74 | 0.65 | 0.61 | 0.63 | 0.70 | 0.67 | 1 | 0.70 | 0.68 | 0.58 | 0.58 | 0.65 | 0.64 | 0.51 | 0.68 | 0.67 | 0.73 | 0.59 | 0.51 | 0.66 | 0.68 | 0.73 |
| MSFT | 0.60 | 0.59 | 0.61 | 0.14 | 0.52 | 0.61 | 0.53 | 0.48 | 0.53 | 0.47 | 0.55 | 0.63 | 0.59 | 0.53 | 0.69 | 0.52 | 0.52 | 0.55 | 0.61 | 0.50 | 0.70 | 1 | 0.62 | 0.57 | 0.51 | 0.57 | 0.56 | 0.41 | 0.57 | 0.53 | 0.59 | 0.51 | 0.34 | 0.61 | 0.58 | 0.61 |
| MXIM | 0.82 | 0.81 | 0.77 | 0.15 | 0.61 | 0.76 | 0.46 | 0.58 | 0.74 | 0.49 | 0.71 | 0.66 | 0.68 | 0.76 | 0.75 | 0.71 | 0.66 | 0.78 | 0.93 | 0.71 | 0.68 | 0.62 | 1 | 0.67 | 0.69 | 0.77 | 0.58 | 0.46 | 0.80 | 0.70 | 0.69 | 0.65 | 0.50 | 0.59 | 0.79 | 0.82 |
| NTAP | 0.65 | 0.68 | 0.72 | 0.16 | 0.70 | 0.72 | 0.46 | 0.65 | 0.64 | 0.42 | 0.70 | 0.55 | 0.64 | 0.66 | 0.63 | 0.67 | 0.65 | 0.66 | 0.66 | 0.55 | 0.58 | 0.57 | 0.67 | 1 | 0.57 | 0.63 | 0.50 | 0.36 | 0.71 | 0.60 | 0.58 | 0.69 | 0.50 | 0.49 | 0.71 | 0.67 |
| NVDA | 0.64 | 0.66 | 0.64 | 0.24 | 0.55 | 0.63 | 0.38 | 0.50 | 0.57 | 0.44 | 0.54 | 0.61 | 0.53 | 0.64 | 0.64 | 0.53 | 0.55 | 0.63 | 0.72 | 0.58 | 0.58 | 0.51 | 0.69 | 0.57 | 1 | 0.63 | 0.48 | 0.36 | 0.69 | 0.58 | 0.52 | 0.52 | 0.37 | 0.46 | 0.60 | 0.68 |
| NVLS | 0.82 | 0.92 | 0.73 | 0.21 | 0.56 | 0.73 | 0.42 | 0.56 | 0.75 | 0.49 | 0.59 | 0.61 | 0.73 | 0.72 | 0.74 | 0.59 | 0.58 | 0.93 | 0.80 | 0.80 | 0.65 | 0.57 | 0.77 | 0.63 | 0.63 | 1 | 0.61 | 0.47 | 0.72 | 0.71 | 0.68 | 0.58 | 0.47 | 0.54 | 0.75 | 0.83 |
| NXTL | 0.61 | 0.57 | 0.59 | 0.25 | 0.55 | 0.65 | 0.43 | 0.46 | 0.61 | 0.52 | 0.57 | 0.52 | 0.68 | 0.65 | 0.59 | 0.55 | 0.57 | 0.57 | 0.61 | 0.61 | 0.64 | 0.56 | 0.58 | 0.50 | 0.48 | 0.61 | 1 | 0.47 | 0.59 | 0.59 | 0.65 | 0.53 | 0.47 | 0.60 | 0.61 | 0.56 |
| PMCS | 0.42 | 0.45 | 0.46 | 0.16 | 0.31 | 0.41 | 0.39 | 0.32 | 0.46 | 0.35 | 0.44 | 0.43 | 0.50 | 0.46 | 0.41 | 0.43 | 0.38 | 0.42 | 0.45 | 0.42 | 0.51 | 0.41 | 0.46 | 0.36 | 0.36 | 0.47 | 0.47 | 1 | 0.43 | 0.42 | 0.48 | 0.33 | 0.25 | 0.48 | 0.42 | 0.48 |
| GLGC | 0.75 | 0.76 | 0.77 | 0.20 | 0.78 | 0.77 | 0.46 | 0.65 | 0.68 | 0.47 | 0.72 | 0.62 | 0.71 | 0.74 | 0.69 | 0.68 | 0.74 | 0.75 | 0.81 | 0.66 | 0.68 | 0.57 | 0.80 | 0.71 | 0.69 | 0.72 | 0.59 | 0.43 | 1 | 0.70 | 0.66 | 0.66 | 0.42 | 0.53 | 0.72 | 0.75 |
| RFMD | 0.72 | 0.67 | 0.73 | 0.24 | 0.59 | 0.72 | 0.43 | 0.62 | 0.75 | 0.37 | 0.64 | 0.52 | 0.68 | 0.67 | 0.66 | 0.64 | 0.64 | 0.70 | 0.73 | 0.67 | 0.67 | 0.53 | 0.70 | 0.60 | 0.58 | 0.71 | 0.59 | 0.42 | 0.70 | 1 | 0.66 | 0.59 | 0.44 | 0.48 | 0.75 | 0.72 |
| SANM | 0.70 | 0.68 | 0.68 | 0.21 | 0.55 | 0.71 | 0.52 | 0.59 | 0.65 | 0.44 | 0.64 | 0.53 | 0.81 | 0.65 | 0.69 | 0.68 | 0.62 | 0.65 | 0.70 | 0.61 | 0.73 | 0.59 | 0.69 | 0.58 | 0.52 | 0.67 | 0.65 | 0.48 | 0.66 | 0.66 | 1 | 0.64 | 0.58 | 0.61 | 0.72 | 0.68 |
| SUNW | 0.64 | 0.66 | 0.65 | 0.21 | 0.66 | 0.67 | 0.52 | 0.58 | 0.67 | 0.44 | 0.76 | 0.56 | 0.67 | 0.67 | 0.70 | 0.68 | 0.62 | 0.62 | 0.66 | 0.55 | 0.59 | 0.51 | 0.65 | 0.69 | 0.52 | 0.58 | 0.53 | 0.33 | 0.66 | 0.59 | 0.64 | 1 | 0.53 | 0.53 | 0.68 | 0.65 |
| TLAB | 0.53 | 0.49 | 0.53 | 0.09 | 0.41 | 0.49 | 0.34 | 0.52 | 0.45 | 0.37 | 0.55 | 0.39 | 0.49 | 0.49 | 0.45 | 0.60 | 0.48 | 0.49 | 0.48 | 0.49 | 0.51 | 0.34 | 0.50 | 0.50 | 0.37 | 0.47 | 0.47 | 0.25 | 0.42 | 0.44 | 0.58 | 0.53 | 1 | 0.42 | 0.54 | 0.48 |
| TMPW | 0.58 | 0.52 | 0.58 | 0.17 | 0.47 | 0.58 | 0.52 | 0.45 | 0.53 | 0.50 | 0.54 | 0.57 | 0.62 | 0.56 | 0.59 | 0.54 | 0.54 | 0.50 | 0.59 | 0.48 | 0.66 | 0.61 | 0.59 | 0.49 | 0.46 | 0.54 | 0.60 | 0.48 | 0.53 | 0.48 | 0.61 | 0.53 | 0.42 | 1 | 0.59 | 0.56 |
| VTSS | 0.78 | 0.74 | 0.87 | 0.15 | 0.64 | 0.82 | 0.47 | 0.69 | 0.75 | 0.44 | 0.72 | 0.57 | 0.71 | 0.74 | 0.72 | 0.73 | 0.70 | 0.76 | 0.78 | 0.72 | 0.68 | 0.58 | 0.79 | 0.71 | 0.60 | 0.75 | 0.61 | 0.42 | 0.72 | 0.75 | 0.72 | 0.68 | 0.54 | 0.59 | 1 | 0.78 |
| XLNX | 0.89 | 0.85 | 0.78 | 0.20 | 0.63 | 0.74 | 0.50 | 0.61 | 0.75 | 0.44 | 0.68 | 0.66 | 0.71 | 0.73 | 0.77 | 0.66 | 0.60 | 0.83 | 0.84 | 0.76 | 0.73 | 0.61 | 0.82 | 0.67 | 0.68 | 0.83 | 0.56 | 0.48 | 0.75 | 0.72 | 0.68 | 0.65 | 0.48 | 0.56 | 0.78 | 1 |

Table 24. Correlation matrix for the stocks which fallen into the 0-th and 7-th clusters.

| | ALTR | AMAT | AMCC | APOL | BRCD | BRCM | CDWC | CIEN | CNXT | CPWR | CSCO | DELL | FLEX | IDTI | INTC | JDSU | JNPR | KLAC | LLTC | MCHP | MOLX | MSFT | MXIM | NTAP | NVDA | NVLS | NXTL | PMCS | GLGC | RFMD | SANM | SUNW | TLAB | TMPW | VTSS | XLNX |
|---|---|---|---|---|---|---|---|---|---|---|---|---|---|---|---|---|---|---|---|---|---|---|---|---|---|---|---|---|---|---|---|---|---|---|---|---|
| ABGX | 0.48 | 0.52 | 0.46 | 0.18 | 0.35 | 0.47 | 0.39 | 0.38 | 0.53 | 0.34 | 0.43 | 0.46 | 0.53 | 0.47 | 0.49 | 0.47 | 0.43 | 0.49 | 0.48 | 0.50 | 0.55 | 0.40 | 0.48 | 0.39 | 0.36 | 0.53 | 0.48 | 0.43 | 0.48 | 0.48 | 0.52 | 0.39 | 0.25 | 0.43 | 0.45 | 0.54 |
| AMGN | 0.29 | 0.37 | 0.33 | 0.14 | 0.13 | 0.22 | 0.28 | 0.23 | 0.37 | 0.23 | 0.30 | 0.30 | 0.32 | 0.26 | 0.35 | 0.33 | 0.26 | 0.37 | 0.35 | 0.27 | 0.41 | 0.28 | 0.37 | 0.27 | 0.21 | 0.34 | 0.29 | 0.30 | 0.27 | 0.29 | 0.32 | 0.24 | 0.21 | 0.34 | 0.28 | 0.37 |
| BGEN | 0.29 | 0.36 | 0.31 | 0.34 | 0.15 | 0.21 | 0.26 | 0.24 | 0.30 | 0.22 | 0.19 | 0.23 | 0.30 | 0.25 | 0.30 | 0.27 | 0.19 | 0.34 | 0.33 | 0.29 | 0.39 | 0.32 | 0.34 | 0.21 | 0.23 | 0.30 | 0.24 | 0.27 | 0.26 | 0.22 | 0.29 | 0.15 | 0.13 | 0.23 | 0.23 | 0.40 |
| CHIR | 0.35 | 0.43 | 0.39 | 0.17 | 0.19 | 0.32 | 0.28 | 0.23 | 0.37 | 0.30 | 0.35 | 0.40 | 0.40 | 0.32 | 0.39 | 0.34 | 0.31 | 0.42 | 0.42 | 0.31 | 0.47 | 0.42 | 0.43 | 0.30 | 0.28 | 0.41 | 0.36 | 0.34 | 0.36 | 0.32 | 0.44 | 0.29 | 0.18 | 0.42 | 0.34 | 0.43 |
| GENZ | 0.37 | 0.38 | 0.39 | 0.15 | 0.23 | 0.32 | 0.30 | 0.33 | 0.33 | 0.35 | 0.26 | 0.31 | 0.39 | 0.32 | 0.33 | 0.32 | 0.30 | 0.39 | 0.33 | 0.38 | 0.45 | 0.34 | 0.38 | 0.33 | 0.31 | 0.41 | 0.31 | 0.32 | 0.35 | 0.33 | 0.34 | 0.24 | 0.16 | 0.41 | 0.36 | 0.38 |
| GILD | 0.31 | 0.39 | 0.34 | 0.16 | 0.22 | 0.33 | 0.31 | 0.25 | 0.37 | 0.26 | 0.31 | 0.33 | 0.46 | 0.32 | 0.31 | 0.37 | 0.33 | 0.37 | 0.37 | 0.37 | 0.40 | 0.26 | 0.38 | 0.26 | 0.31 | 0.39 | 0.42 | 0.29 | 0.35 | 0.34 | 0.42 | 0.23 | 0.18 | 0.40 | 0.32 | 0.35 |
| HGSI | 0.48 | 0.57 | 0.54 | 0.10 | 0.40 | 0.49 | 0.44 | 0.40 | 0.55 | 0.39 | 0.48 | 0.48 | 0.55 | 0.53 | 0.52 | 0.48 | 0.43 | 0.53 | 0.53 | 0.49 | 0.55 | 0.44 | 0.54 | 0.42 | 0.41 | 0.53 | 0.50 | 0.49 | 0.50 | 0.50 | 0.51 | 0.42 | 0.29 | 0.45 | 0.47 | 0.57 |
| IDPH | 0.31 | 0.33 | 0.35 | 0.11 | 0.18 | 0.29 | 0.25 | 0.23 | 0.32 | 0.22 | 0.31 | 0.31 | 0.37 | 0.30 | 0.27 | 0.30 | 0.31 | 0.32 | 0.31 | 0.32 | 0.36 | 0.25 | 0.34 | 0.25 | 0.22 | 0.34 | 0.30 | 0.30 | 0.36 | 0.29 | 0.39 | 0.21 | 0.16 | 0.38 | 0.29 | 0.36 |
| IMCL | 0.26 | 0.30 | 0.30 | 0.18 | 0.22 | 0.27 | 0.28 | 0.21 | 0.30 | 0.27 | 0.30 | 0.32 | 0.39 | 0.29 | 0.27 | 0.34 | 0.30 | 0.30 | 0.30 | 0.27 | 0.37 | 0.26 | 0.31 | 0.25 | 0.24 | 0.30 | 0.37 | 0.25 | 0.32 | 0.31 | 0.33 | 0.23 | 0.21 | 0.39 | 0.26 | 0.28 |
| MEDI | 0.47 | 0.52 | 0.50 | 0.23 | 0.31 | 0.42 | 0.33 | 0.38 | 0.51 | 0.38 | 0.44 | 0.42 | 0.49 | 0.38 | 0.48 | 0.46 | 0.42 | 0.48 | 0.48 | 0.44 | 0.61 | 0.48 | 0.49 | 0.36 | 0.36 | 0.51 | 0.45 | 0.44 | 0.47 | 0.40 | 0.46 | 0.36 | 0.27 | 0.48 | 0.46 | 0.53 |
| MLNM | 0.54 | 0.54 | 0.53 | 0.21 | 0.43 | 0.52 | 0.46 | 0.46 | 0.55 | 0.44 | 0.54 | 0.51 | 0.58 | 0.50 | 0.51 | 0.53 | 0.48 | 0.53 | 0.51 | 0.52 | 0.60 | 0.50 | 0.54 | 0.45 | 0.35 | 0.54 | 0.58 | 0.49 | 0.53 | 0.50 | 0.56 | 0.45 | 0.35 | 0.52 | 0.48 | 0.56 |
| PDLI | 0.42 | 0.45 | 0.46 | 0.16 | 0.31 | 0.41 | 0.39 | 0.32 | 0.46 | 0.35 | 0.44 | 0.43 | 0.50 | 0.46 | 0.41 | 0.43 | 0.38 | 0.42 | 0.45 | 0.42 | 0.51 | 0.41 | 0.46 | 0.36 | 0.36 | 0.47 | 0.47 | 0.39 | 0.43 | 0.42 | 0.48 | 0.33 | 0.25 | 0.48 | 0.42 | 0.48 |

In Table 24 correlation coefficients between the companies, which have fallen into the darkest blue and reddest clusters (the distance between these clusters is maximum) are given. The mean value and the standard deviation of the correlation matrix for the companies forming NASDAQ100 are equal to 0.47 and 0.18 correspondingly. Mean value of correlation coefficients form Table 24, is equal to 0.37. As one would expect, average value of correlation coefficients between companies, which have fallen into clusters №0 and №7 is less then the average of all coefficients of correlations.

# 8    Conclusion

The paper shows advantages of self-organizing maps in comparison with traditional models of statistical data analysis. In particular we compare the clusterization of DJIA and NASDAQ100 portfolios by hierarchical trees and SOM methods. It was shown, that:

1. In the case of DJIA the results of both methods appeared quite similar. It means that ultra metric methods practically do not lead to an error in clusterization on a small data sample.
2. The clusterization of NASDAQ100 shows that the results of SOM are essentially subtle than methods of ultrametrics. Such results mean, in particular, that conventional statistical methods are imperfect in application to the large data samples.

Thus one can conclude that the SOM method is more relevant to the problems where processing of the large data samples is required. In particular, this method can be used for the forming and dynamical management a well diversificated portfolio of stocks.